\documentclass[a4paper]{article}

\usepackage[pages=all, color=black, position={current page.south}, placement=bottom, scale=1, opacity=1, vshift=5mm]{background}

\usepackage[margin=1in]{geometry} 

\usepackage{IEEEtrantools}
\usepackage{cite}
\usepackage{amsmath,amssymb,amsfonts}
\usepackage{algorithmic}
\usepackage{algorithm}
\usepackage{lipsum}

\usepackage{graphicx}
\usepackage{epsfig}
\usepackage{amsmath}
\usepackage{bbm}
\usepackage{booktabs}
\usepackage{multirow}
\usepackage{tabularx}
\usepackage[normalem]{ulem}
\usepackage{makecell}

\usepackage{siunitx}
    \usepackage[caption=false]{subfig}
\usepackage{textcomp}
\usepackage{xcolor}
\usepackage{verbatim}

\DeclareMathOperator*{\maximize}{maximize}
\def\BibTeX{{\rm B\kern-.05em{\sc i\kern-.025em b}\kern-.08em
    T\kern-.1667em\lower.7ex\hbox{E}\kern-.125emX}}
\usepackage[nopostdot,nogroupskip,style=super,nonumberlist,acronym]{glossaries}
\newacronym{aoi}{AoI}{Age of Information}
\newacronym{aodv}{AODV}{Ad hoc On-demand Distance Vector}
\newacronym{cnn}{CNN}{convolutional neural networks}
\newacronym{deepl}{DL}{Deep Learning}
\newacronym{dod}{DoD}{depth of discharge}
\newacronym{dqn}{DQN}{deep Q-learning}
\newacronym{gsl}{GSL}{ground-to-satellite link}
\newacronym{isl}{ISL}{inter-satellite link}
\newacronym{leo}{LEO}{low Earth orbit} 
\newacronym{ml}{ML}{Machine Learning}
\newacronym{mdp}{MDP}{Markov decision process}
\newacronym{ngeo}{NGSO}{Non-geostationary orbit}
\newacronym{ngso}{NGSO}{Non-geostationary orbit}
\newacronym{olsr}{OLSR}{optimized link state routing protocol}
\newacronym{ospf}{OSPF}{Open Shortest Path First}
\newacronym{pan}{PAN}{Path-Aware Networking}
\newacronym{qos}{QoS}{Quality of Service}
\newacronym{rl}{RL}{Reinforcement Learning}
\newacronym{drl}{DRL}{Deep \gls{rl}}
\newacronym{dnn}{DNN}{Deep Neural Network}
\newacronym{dql}{DQL}{Deep Q-learning}
\newacronym{ql}{QL}{Q-learning}
\newacronym{e2e}{E2E}{end-to-end}
\newacronym{bgp}{BGP}{Border Gateway Protocol}
\newacronym{ibgp}{iBGP}{interior Border Gateway Protocol}
\newacronym{ebgp}{eBGP}{exterior Border Gateway Protocol}
\newacronym{as}{AS}{Autonomous System}
\newacronym{relu}{ReLu}{Rectified Linear Unit}
\newacronym{cdf}{CDF}{Cumulative Distribution Function}
\newacronym{ue}{UE}{User Equipment}
\newacronym{gps}{GPS}{Global Positioning System}
\newacronym{pomdp}{POMDP}{Partially Observable \gls{mdp}}
\newacronym{snr}{SNR}{signal-to-noise ratio}
\newacronym{awgn}{AWGN}{additive-white Gaussian noise}
\newacronym{bm}{BM}{benchmark}
\newacronym{fifo}{FIFO}{first-in first-out}
\newacronym{ip}{IP}{Internet Protocol}
\newacronym{gsd}{GSD}{Ground Sample Distance}
\newacronym{eo}{EO}{Earth Observation}
\newacronym{iot}{IoT}{Internet of Things}
\newacronym{semcom}{SemCom}{Semantic Communications}
\newacronym{ssim}{SSIM}{structural similarity index measure}
\newacronym{sdg}{SDG}{Sustainable Development Goal}
\newacronym{ai}{AI}{Artificial Intelligence}
\newacronym{mse}{MSE}{Mean Square Error}
\newacronym{fid}{FID}{Fréchet Inception Distance}
\newacronym{jscc}{JSCC}{Joint Source and Channel Coding}
\newacronym{kg}{KG}{Knowledge Graph}
\newacronym{gtfp}{GTFP}{ground track frame period}
\newacronym{gs}{GS}{ground station}
\newacronym{fov}{FOV}{Field of View}
\newacronym{fso}{FSO}{free-space optical}
\newacronym{ntn}{NTN}{Non-Terrestrial Networks}
\newacronym{lsatc}{LSatC}{Low Earth Orbit Satellite Constellations}
\newacronym{fm}{FM}{Foundation Model}
\newacronym{genai}{GenAI}{Generative AI}
\newacronym{ul}{UL}{uplink}
\newacronym{dl}{DL}{downlink}
\newacronym{nir}{NIR}{near infrared}
\newacronym{uv}{UV}{ultraviolet}
\newacronym{fps}{FPS}{frames per second}
\newacronym{kb}{KB}{Knowledge Base}
\newacronym{soa}{SoA}{State-of-the-Art}
\newacronym{cpu}{CPU}{Central Processing Unit}
\newacronym{rf}{RF}{Radio Frequency}
\newacronym{gpu}{GPU}{Graphics Processing Unit}
\newacronym{vnir}{VNIR}{Visible and Near Infra-Red}
\newacronym{swir}{SWIR}{Short Wave Infra-Red}
\newacronym{mwir}{MWIR}{Mid Wave Infra-Red}
\newacronym{lwir}{LWIR}{Long Wave Infra-Red}
\newacronym{aeos}{AEOS}{agile EO satellite}
\newacronym{aeossp}{AEOSSP}{AEOS scheduling problem}
\newacronym{vtw}{VTW}{visible time window}
\newacronym{otw}{OTW}{observation time window}
\newacronym{sth}{STH}{scheduling time horizon}
\newacronym{map}{mAP}{mean Average Precision}
\newacronym{iou}{IoU}{Intersection over Union}
\newacronym{flop}{FLOP}{floating-point operation}
\newacronym{rv}{RV}{random variable}
\newacronym{kpi}{KPI}{Key Performance Indicator}
\newacronym{ga}{GA}{genetic algorithm}
\newacronym{nms}{NMS}{Non-Maximum Suppression}
\newacronym{clt}{CLT}{Central Limit Theorem}
\newacronym{bsp}{BSP}{Bulk-Synchronous Parallel}
\DeclareMathOperator*{\minimize}{minimize}
\newcommand{\set}{\mathcal}

\definecolor{darkgreen}{rgb}{0.0, 0.5, 0.0}
\newcommand{\ammm}[1]{\textcolor{darkgreen}{#1}}

\usepackage{comment}

\title{{Edge Intelligence for Satellite-based Earth Observation: Scheduling Image Acquisition and Processing}}
\author{Beatriz Soret, Antonio M. Mercado-Martínez, Antonio Jurado-Navas, \\Nicolai D. Lyholm, Marco Moretti, Petar Popovski, and Israel Leyva-Mayorga}
\vspace{-0.4cm}
\begin{document}
\maketitle
\renewcommand{\thefootnote}{}
\footnotetext{B. Soret, A. M. Mercado-Martínez, and A. Jurado-Navas are with the Telecommunications Research Institute (TELMA), Universidad de Málaga, Málaga, Spain. N. D. Lyholm, P. Popovski and I. Leyva-Mayorga are with the Connectivity Section, Aalborg University, Aalborg, Denmark. Marco Moretti is with the University of Pisa, Pisa, Italy and also with the National Inter-University Consortium for Telecommunications (CNIT), 43124 Parma, Italy. This work was partially funded by ESA SatNEx V (prime contract no. 4000130962/20/NL/NL/FE). The view expressed herein can in no way be taken to reflect the official opinion of the European Space Agency.
The work of B. Soret, A.M. Mercado-Martínez and A. Jurado-Navas is further supported by the Spanish Ministerio de Ciencia e Innovación under grant PID2022-136269OB-I00 funded by MCIN/AEI/10.13039/501100011033 and “ERDF A way of making Europe”. The work of A.M. Mercado-Martínez is also supported by Grant DGP\_PRED\_2024\_01603, funded by the Consejería de Universidad, Investigación e Innovación of Junta de Andalucía and the European Union. The work of Petar Popovski and Israel Leyva-Mayorga is partially supported by the Velux Foundation, Denmark, through the Villum Investigator Grant WATER, nr. 37793. The work of Marco Moretti is  also partially supported by the Italian Ministry of Education and Research (MUR) in the Framework of the FoReLab Project (Departments of Excellence).
}

\begin{abstract}
Modern \gls{eo} missions generate massive volumes of imagery that challenge existing downlink and ground-processing capabilities, particularly for time-critical applications. This work investigates how a \gls{leo} satellite constellation equipped with heterogeneous edge computing resources can enable real-time semantic processing of data acquired by \gls{eo} satellites. We introduce an energy-aware framework that optimizes the use of resources accounting for data acquisition, computing, and communication constraints. Although we focus on maritime surveillance, the formulation is task-agnostic and accommodates a broad class of semantic and goal-oriented inference problems.
Specifically, we formulate two coupled optimization problems: (i) observation scheduling, which selects image acquisition opportunities while accounting for turbulence-induced image degradation and energy budget, and (ii) processing scheduling, which allocates semantic workloads across onboard and ground processors. We evaluate these mechanisms for the task of detection and localization of vessels, for which we quantify the benefits of turbulence-aware observation scheduling for preserving image quality and experimentally characterize the execution-time distribution of YOLOv8 on different computing platforms. Results demonstrate that task- and turbulence- aware observation scheduling can significantly improve the quality and quantity of observed targets. Furthermore, cooperative edge processing within the constellation substantially reduces power consumption compared to traditional downlink-centric architectures. These findings highlight the potential of distributed edge intelligence to enhance the responsiveness and autonomy of future satellite-based \gls{eo} systems.

\vspace{-0.4cm}

\end{abstract}

 \glsresetall
\section{Introduction}

We are witnessing a surge in satellite-based \gls{eo}~\cite{eobook}, driven by scientific, commercial, and geopolitical factors, with the number of satellites expected to triple by 2033~\cite{novaspace2024_EOsatellites}. This growth reflects the essential role of \gls{eo} in providing critical information for climate and environmental monitoring~\cite{yang2013climatechange}\cite{persello2022SDGs}, maritime surveillance~\cite{soldi2021maritime}, disaster management~\cite{oddo2019flood}, and beyond.
Historically, \gls{eo} applications have downloaded vast amounts of collected raw, or minimally processed, data to a server on ground, to be stored and processed at some point in the future. 

This is changed radically by making the EO data part of real-time computation-communication loops within the emerging trend towards on-board intelligence. There are several drivers of this trend:

\noindent\emph{(i)} Satellite and space technologies are drastically reducing the cost of the missions through specialized yet affordable spacecrafts, leading to a densification of \gls{leo} satellite networks and increasing the volume of \gls{eo} data generated in orbit.

\noindent\emph{(ii)} Edge computing enables the use of the computing capabilities of the network infrastructure to analyze information closer to the source, resulting in reduced latency and bandwidth. In the case of space, \gls{leo} satellite edge nodes can process raw \gls{eo} data before being sent to the ground, de-congesting the communication network, specifically the feeder link connecting the space segment to the ground segment~\cite{Ley23TCOM}. 

\noindent\emph{(iii)} \gls{ai} and Deep Learning are driving remarkable progress in data processing, including image and video, by transforming raw data into actionable intelligence. When combined with \emph{\gls{semcom}}~\cite{Luo2022semanticmagazine} principles, these methods extract and transmit the informative content (meaning, relevance and mission value), which allows a great reduction in data volume~\cite{semanticEO}. Although semantic compression and feature extraction have already been adopted in space missions, such as ESA's Sentinel-2, they typically operate in isolation at the source, without considering the end-to-end communication context or task-specific objectives. In contrast, the \gls{semcom} concept is a broader paradigm that envisions the joint optimization of sensing, processing, transmission, and interpretation, guided by a shared \gls{kb} and aligned with the semantic goals of the application at the intelligent communicating nodes~\cite{deniz2023semantic}.  
\begin{figure}[t]
\centering
\includegraphics[width=3.5 in] {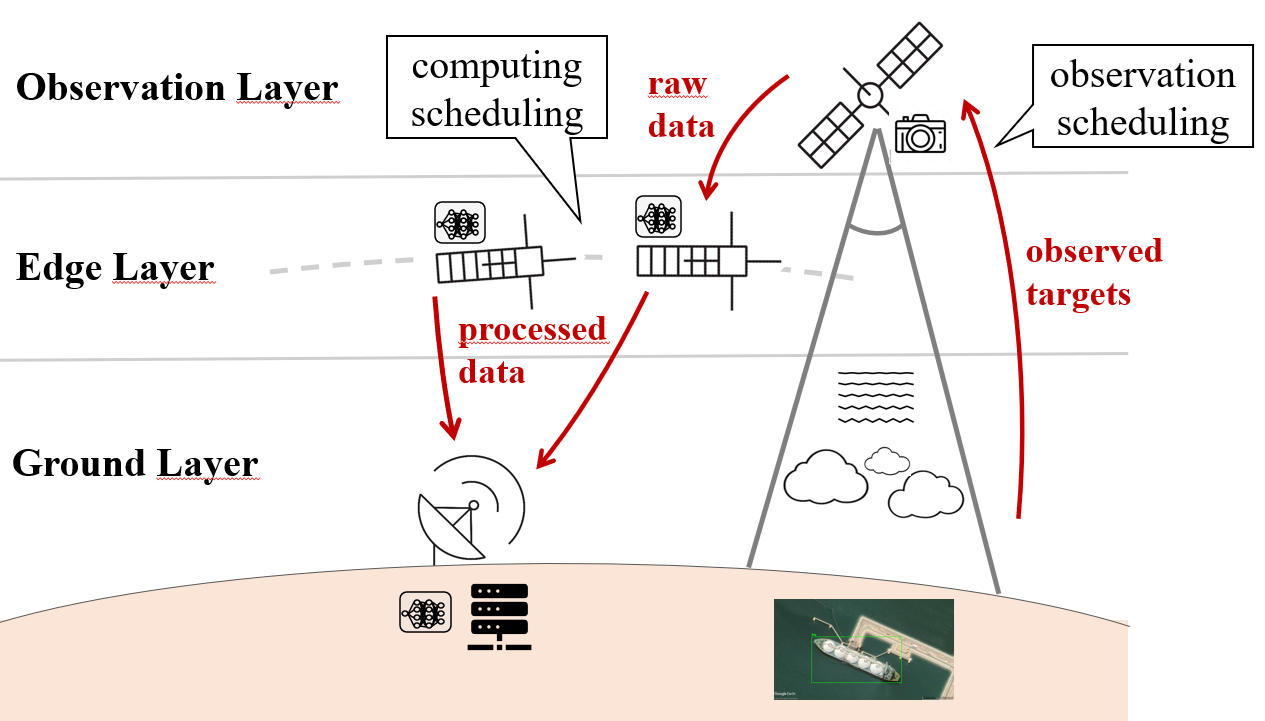}
\caption{Scenario overview: \gls{eo} satellites capture images of a set of tracked objects (vessels), with image quality affected by atmospheric turbulence. Raw images are transmitted to the edge layer via \glspl{isl} for semantic compression (vessel detection and localization), and processed data transmitted to ground via \gls{dl}. Both the \gls{eo} satellites and the edge constellation operate within the \gls{leo} layer. 
\vspace{-0.5cm}} \label{fig:scenario}
\end{figure}

Together, these developments form the foundation for edge intelligence in \gls{eo}~\cite{ruan2025edgeEO, denby2019arch}, where satellites do not simply observe the Earth, but understand it, enabling dynamic, efficient, and autonomous sensing and processing, considering the final goal of the \gls{eo} task. In this paper, we focus on two core procedures for this vision: (1) {Scheduling of observations}, which determines what to observe, when, by which satellites, and under which acquisition parameters. The latter includes adjusting the attitude control subsystem at the spacecrafts for dynamic adjustment of pointing,  accounting for impairments caused by the atmospheric turbulence. (2) {Scheduling of processing}, which allocates data processing and communication tasks between satellite edge nodes and ground resources. These optimizations explicitly consider constraints in terms of time, energy, as well as computation and communication resources.

Our proposed scenario integrating {observation and computation} scheduling is illustrated in Fig.~\ref{fig:scenario}: A set of \gls{eo} satellites capture images subject to atmospheric turbulence, and may belong to, or operate independently from, the \gls{leo} constellation providing edge computing capabilities. Although the theoretical framework is general and task-agnostic to accommodate a broad class of semantic and goal-oriented inference problems, we focus on the \gls{eo} task of observing a set of vessels (the targets) on the Earth's surface. The resource orchestrator at the \gls{gs} translates this query into an \emph{{observation} scheduling}, to decide what to be observed and the corresponding observation parameters, and a \emph{processing scheduling}, to decide how image processing is distributed between satellite edge nodes and, if necessary, ground resources. The scheduled satellites run an object detection and localization algorithm on the acquired images and encode the result. The processed information is then transmitted to the ground for final aggregation, interpretation, and delivery to the end user. 

Extensive research has addressed the data acquisition process, including agile satellites, as reviewed in Section~\ref{sec:eo}. However, despite being a mature technology, several aspects remain unexplored for real-time, quality-aware optimization, with a few exceptions~ \cite{mercadomartinez2025schedulingagileearthobservation}. The second component of interest, edge computing in space, has only recently begun to receive attention. Relevant use cases include mobile edge computing, \gls{iot}, \gls{eo} and federated learning~\cite{YIN2025103316, renchao2020satedgecomputing, bho2024FL, denby2019arch, Denby2023, ruan2025edgeEO}. An example of a joint observation and communication optimization is found in \cite{jointobservationcomputation}. However, the observation model does not capture image quality or atmospheric turbulence, and the processing model also includes simplifications that moves it away from the optimal allocation.

The rest of the paper is organized as follows. In Section \ref{sec:eo} we describe the fundamentals of \gls{eo}, including the key elements of the considered optimizations: the data acquisition and observation scheduling, the meteorological and atmospheric turbulences, and the data processing techniques. Then, the system model is introduced in Section \ref{sec:systemmodel}. In Section \ref{sec:optimization}, the optimization problems for the observation scheduling and the edge computing are formulated, and Section~\ref{sec:results} includes the evaluation results that demonstrate the achieved gains. Section~\ref{sec:conclusions} concludes the paper.

\section{Fundamentals of Earth Observation} \label{sec:eo}

 \gls{eo} satellites generate large amounts of  data about the Earth's surface, water bodies, and atmosphere. For example, the ESA Sentinel missions acquire approximately 12 TB of images daily, while NASA missions collectively provide \mbox{20-30 TB} per day. These volumes are rapidly increasing as new missions are planned and launched. \gls{eo} images are used for different purposes requiring computer vision processing and inference, e.g., object localization and detection, semantic labeling of image regions, change detection, or time series analysis. Table ~\ref{tab:soa} provides a representative list of \gls{eo} missions and \gls{soa} processing algorithms. 

\subsection{Data acquisition and scheduling}  
The \gls{eo} satellite payload includes various remote sensing instruments, such as cameras, radar, spectrometers, or thermal sensors. Optical cameras are often multi-spectral, ranging from the \gls{nir} to the \gls{uv} wavelengths. A full frame is typically obtained by superimposing frames captured at different wavelengths. For example, a color image is produced by combining red, green, and blue (RGB) channels, whereas a \emph{panchromatic} image is obtained from a wide-band capture that provides higher spatial resolution.

The area covered and the quality of the resulting images are primarily affected by \emph{(i)} the altitude of the orbit; \emph{(ii)} the \gls{fov}, which is the angular width that determines the extent of the observable area captured by the camera sensor; \emph{(iii)} the \gls{gsd}, which represents the actual distance on the ground between the centers of adjacent pixels, determining the spatial resolution of the imagery~\cite{Ley23TCOM}. Typically, lower altitudes correspond to lower \gls{gsd} and \gls{fov} values. Since high-resolution imagery requires low \gls{gsd} values, \gls{leo} satellites are the prevalent choice for \gls{eo}. As the captured frame extends further off-nadir (away from the point directly below the satellite), the \gls{gsd} increases, degrading image spatial resolution \cite{NAG2018891}. Therefore, accurate control of the observed surface footprint is essential for high-resolution \gls{eo}. Modern attitude control subsystems enable dynamic adjustment of pointing to meet such observation requirements. Additionally, \emph{pansharpening}, where a multispectral frame is combined with a panchromatic frame, can be used to reduce the \gls{gsd} and enhance image quality.

The new generation of \glspl{aeos}, equipped with full three-axis attitude control (roll, pitch, and yaw)~\cite{Wang_2021}, significantly enhances data acquisition capabilities, extending the duration of \glspl{vtw}, while enabling multiple \glspl{otw} for a given observation target within a single orbital pass. We refer to \gls{vtw} as the time interval during which a satellite has visibility of a target, whereas \glspl{otw} denote the specific subintervals during which the actual imaging takes place. Although this improved flexibility expands the range of feasible observation schedules, it also complicates the \gls{otw} selection when multiple targets are involved in the same \gls{sth}. This challenge is known as \gls{aeossp}, which consists of determining the target observation sequence, the corresponding \gls{otw}, and, in multi-satellite scenarios, the satellite responsible for executing such actions \cite{alma991001513474404886}. The objective is to maximize the observation profit while satisfying a set of operational and resource constraints. The observation profit is a metric that quantifies the value of an observation and depends on the purpose of the mission; common definitions include image quality \cite{PENG201984}, target priority \cite{10058020}, or the influence of atmospheric conditions \cite{cloud}. Given the NP-hard nature of the \gls{aeossp} \cite{LEMAITRE2002367}, where computational complexity grows rapidly with the number of satellites and targets, exact optimization approaches become impractical for large problem instances. Consequently, a large volume of literature has focused on heuristic and learning-based methods \cite{mercadomartinez2025schedulingagileearthobservation, 10835124}, especially when resources are limited and real-time scheduling is required. Recent research integrates the \gls{aeossp} into the operation of the entire satellite system to jointly optimize observation, computation, and communication tasks \cite{jointobservationcomputation, SHANG202592, ROCHA2025107212}.
\begin{table*}[t]
\caption{Earth Observation (EO) missions and \gls{soa} image processing algorithms.}

\centering
\renewcommand{\arraystretch}{1.2}
\begin{tabular}{@{} p{2cm} p{3cm} p{4.7cm} p{4.7cm}@{}}
\toprule
\multicolumn{4}{c}{\textbf{EO missions}}\\
\midrule
\textbf{Name}& \textbf{Agency/Company}& \textbf{Constellation} & \textbf{Imaging capabilities}\\\midrule
Sentinel-2 & ESA & 2  satellites at 786 km. &13 spectral bands from \gls{vnir} to \gls{swir}.\\
GOES & NOAA and NASA & 3 geostationary satellites. & 16 spectral bands: \gls{vnir}, \gls{mwir}, and \gls{lwir}. \\
PlanetScope \cite{PlanetScope} & PlanetLabs & $>430$ 3U CubeSats Doves and SuperDoves at 475-525 km. & 8-band multispectral images. \\
WorldView Series \cite{worldview-13} 
  & Maxar & 3  satellites at 496 km (WorldView-1), 770 km (WorldView-2) and 617 km (WorldView-3). &  WorldView-1: panchromatic images. WorldView-2 and 3: 8-band multispectral  images.\\
  \midrule
  \multicolumn{4}{c}{\textbf{Image processing algorithms}}\\
  \midrule
  \textbf{Name}& \textbf{Application}& \textbf{Architecture and characteristics} & \textbf{Performance}\\\midrule
  JPEG2000 & Compression & Scalable lossy and lossless compression. & --\\
  Single Shot Detector (SSD) 
& Object detection and localization& Single CNN to predict multiple bounding boxes and their class probabilities from multiple feature maps. & \Gls{map} @0.5 $\geq 68\%$ on PASCAL VOC2007 dataset.\\
YOLO (v8) \cite{yolov8_ultralytics} 
&  Classification, segmentation, object detection, and tracking& Multiple CNN and fully connected layers to predict the bounding boxes of the objects & \gls{map} @0.5 = $76.8\%$.\\
GOTURN \cite{held2016learningtrack100fps} & Time series analysis& CaffeNet architecture to predict the bounding box coordinates of an object in the frame from a previous one. & Accuracy = 61\%, Robustness = 90\%.\\
EndNet \cite{Hong_EndNet_2022} & Multimodal data fusion& Encoder-decoder NN & Averaged accuracy = $93.88\%$.\\
\bottomrule
\end{tabular}

\label{tab:soa}

\end{table*}

\begin{figure*}[t]
\centering
\includegraphics[scale=0.30]{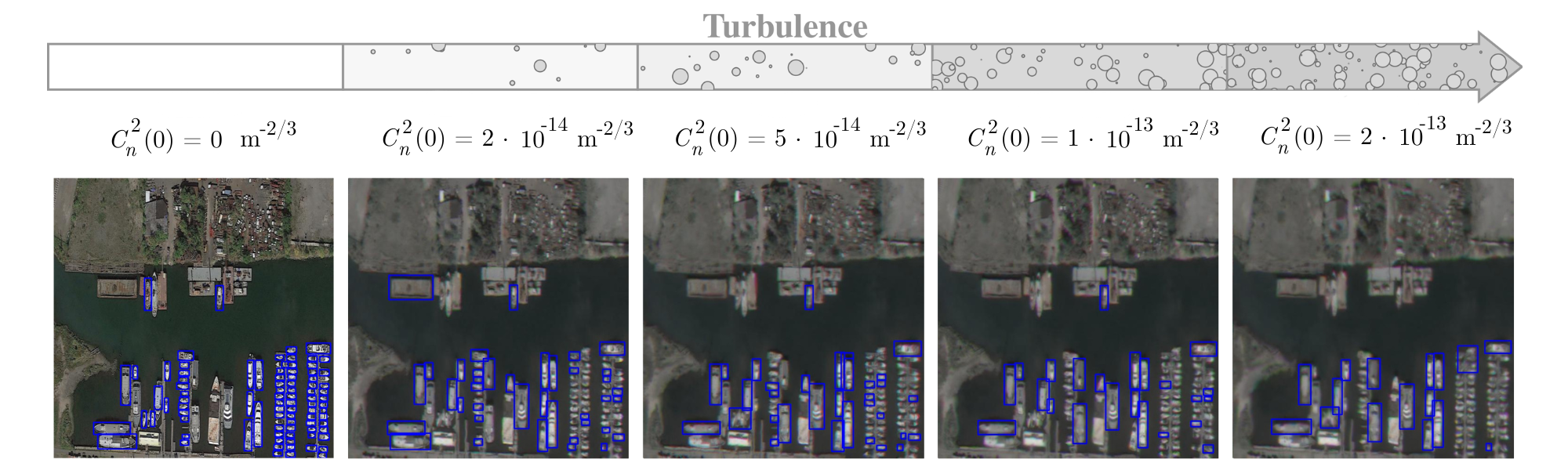}
\caption{Impact of the atmospheric turbulence in ship detection and localization using YOLOv8. The severity of the atmospheric turbulence is parametrized in the structure parameter $C_n^2(0)$ \cite{andrews2005laser}. The input images and mission parameters are from \cite{ships-google-earth_dataset}. The output of the algorithm are the blue bounding boxes that include coordinates and confidence about the detection.}
\vspace{-0.4 cm}\label{fig:atmospheric}
\end{figure*}
\begin{figure*}[h]
\centering
\includegraphics[scale=0.3]{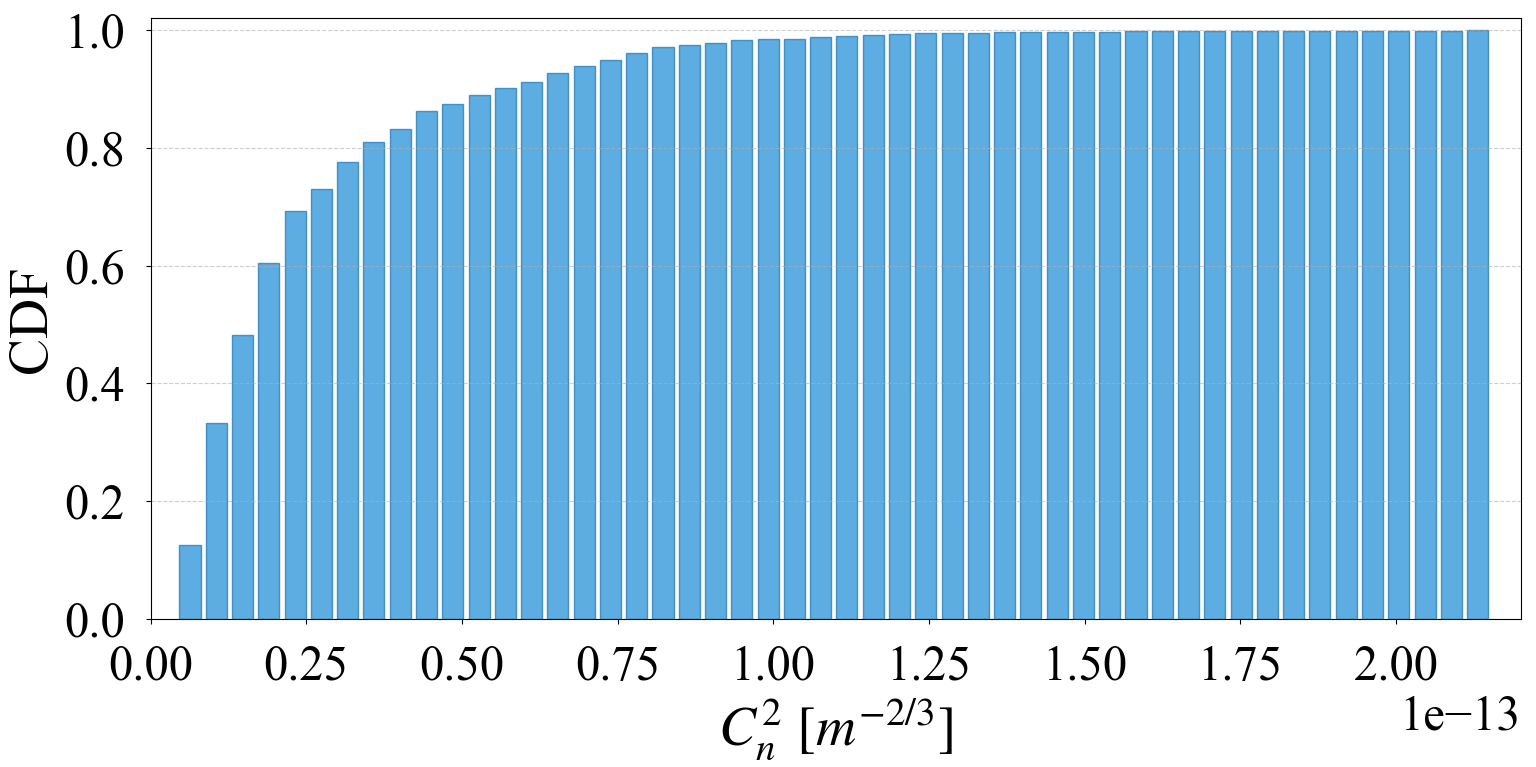}
\caption{\gls{cdf} for $C_n^2(0)$ based on experimental ground-level measurements. Measures taken from \cite{Jurado-Navas:12}.}
\vspace{-0.4 cm}\label{fig:cn2_cdf}
\end{figure*}
\begin{table*}[!t]
\caption{Overview of $\mathbf{C_n^2}$ estimation techniques .}
\centering
\renewcommand{\arraystretch}{1.2}
\begin{tabular}{@{} p{2cm} p{3cm} p{4.7cm} p{4.7cm}@{}}
\toprule
\multicolumn{4}{c}{\textbf{Comparison of $\mathbf{C_n^2}$ Estimation Techniques and Associated References}}\\
\midrule
\textbf{Category}& \textbf{Technique}& \textbf{Method Basis} & \textbf{Key Limitation / Challenge}\\\midrule
 & Speckle-based Observation (ML) \cite{Ciarella2025}  & ML reconstruction of the ${C_n^2}$ profile from single-shot star/beacon speckle patterns. &Lack of a clear analytical theory connecting speckle observations to the turbulence profile. For GSs only.\\
Passive Optical & Image Gradient Method \cite{Saha:22,McCrae:17} & Calculates ${C_n^2}$ from the statistics of image shifts (tilt variance) in long-range video, simplifying physics-based assumptions. & Influenced by turbulence near the camera, gradient implementation and ROI choice. Critical for implementing and processing on-board a satellite.\\
 & Physics-based CNN (Hybrid) \cite{Saha:22, Ciarella2025} & Integrates convolutional layers with a differentiable gradient approach, explicitly incorporating optical parameters like aperture ($D$) and distance ($L$). & Suffer from poor generalization across unseen datasets or changes in optical parameters. \\
\hline
  & Tatarskii Gradient (NWP/Satellite) \cite{Meier:14,Fiorino:14,Meier:15,Wang:20,Frehlich:10} & Calculates ${C_n^2}$ using the vertical gradient of potential temperature derived from satellite (AIRS) or Numerical Weather Prediction (NWP) data. & Requires empirical estimation of the outer length scale. Suffers from  zero-gradient problem in neutrally-buoyant boundary layer.\\
Meteorological & Horizontal Structure Functions \cite{Frehlich:10,Wang:20}& Estimates ${C_n^2}$ from horizontal spatial statistics from NWP, applying corrections for inherent model smoothing. & Produces area-averaged estimates that fail to capture local, highly intermittent turbulence variability. \\
    & Second-Order Turbulence Closure \cite{Wang:20}& Predictive approach solving prognostic equations for scalar variances 
    within NWP turbulence closure schemes. Turbulent mixing effects. & Largest errors near the top of the boundary layer. Requires accurate determination of Convective Boundary Layer  (CBL) height ($z_i$). \\
\hline
Instrumental & Scintillometers \cite{Fiorino:14,McCrae:17,Saha:22} & Double-ended optical measurement of irradiance scintillation and angle-of-arrival fluctuations. Measures path-averaged ${C_n^2}$. & Bulky, costly, and requires precise alignment over long distances. Systematic errors due to poor focal alignment. For communications with GSs. \\
& Thermosondes\cite{Ciarella2025,Frehlich:10,Meier:15,McCrae:17} & Balloon-borne fine-wire probes measuring temperature turbulence (${C_T^2}$) directly over a wide range of elevations. & Estimation error is dominated by the intrinsic intermittency of turbulence (RMS error $\approx 50\%$). Intrusive deployment. \\
\bottomrule
\end{tabular}

\label{tab:cn2}
\end{table*}
\subsection{Meteorological effects}

Meteorological factors play a key role in selecting the most suitable image acquisition and processing techniques, directly impacting their effectiveness and accuracy. The atmosphere acts as a massive thermodynamic system  
with slow, periodic oscillations caused by atmospheric tides, and small fluctuations in the atmospheric refractive index due to slight temperature variations. 

The thermal micro-variations are explained by the appearance of convective air currents. These are caused, horizontally, by the progressive warming of the ground by the Sun and, vertically, by velocity gradients between different atmospheric layers. The latter modifies the uniform characteristics of medium's viscosity due to dynamic mixing and random subflows, called turbulent vortices or eddies \cite{andrews2005laser},
which induce characteristic image aberrations, compromising the achievable angular camera resolution.  

Atmospheric turbulence can be modelled through phase screens, which contain the phase fluctuation related to the refractive index fluctuation spectrum. To realistically simulate the way lightwave aberrations are captured by a camera, a thin phase screen technique is employed, based on the Kolmogorov power spectrum density of the refractive index~\cite{andrews2005laser}. This phase screen is then superimposed on the wavefront of the light as it propagates through the simulated atmospheric layer. Figure \ref{fig:atmospheric} shows an example of the impact of the atmospheric turbulence in the performance of the state-of-the-art object detection and localization algorithm, YOLOv8, applied to a pre-trained \gls{eo} database of images collected from a $500$ m eye altitude level~\cite{ships-google-earth_dataset}. As the turbulence becomes more severe, parameterized in larger refractive index fluctuations and structure parameter $C_n^2$ \cite{andrews2005laser}, the algorithm struggles to detect vessels. However, this effect is often overlooked in system optimization.
Fig.~\ref{fig:cn2_cdf} shows an empirical \gls{cdf} of this parameter as derived from experimental ground-level measurements \cite{Jurado-Navas:12}.

\subsubsection{Atmospheric channel estimation}
Accurate estimation of the refractive index structure constant, $C_n^2$, is essential for modeling optical propagation, outage probability, scintillation, and beam-wander effects in satellite-to-ground optical links. 
Unlike ground-based astronomy or terrestrial FSO, where $C_n^2$ can be measured directly using SCIDAR (Scintillation Detection and Ranging), MASS (Multi Aperture Scintillation Sensor), DIMM (Differential Image Motion Monitor), or scintillometers, LEO platforms impose strict constraints on payload mass, power, pointing, and telemetry budget. Consequently, a rigorous assessment of all estimation methodologies is required to identify those suitable for spaceborne implementation.

In Table~\ref{tab:cn2} we summarize the most robust and practical methods for estimating $C_n^2$ in LEO missions based on physical arguments, algorithmic feasibility, and published empirical evidence. Passive optical approaches, including speckle-based, image-gradient, and hybrid CNN–physics methods, are fundamentally unsuitable for spaceborne implementation, as they are primarily sensitive to turbulence local to the receiver, which is negligible at satellite altitude, while the dominant contribution to optical degradation arises in the lower atmosphere; moreover, their assumptions of quasi-stationary, long-path geometries are incompatible with rapid orbital motion and impose significant on-board computational and calibration burdens. Instrumental profilers such as scintillometers, SCIDAR, MASS, and radiosondes are likewise infeasible for LEO platforms and remain limited to ground-based validation. In contrast, meteorological estimation methods based on Numerical Weather Prediction (NWP) models and satellite sounders provide a physically consistent and operationally viable solution, offering global, continuous, and forecastable $C_n^2$ profiles with full four-dimensional coverage and inherent sensitivity to the atmospheric layers that dominate optical propagation. These characteristics enable seamless integration with link-budget analyses and end-to-end simulators without imposing additional payload, power, or calibration requirements on the spacecraft.
While existing SemCom frameworks increasingly account for physical and semantic noise, atmospheric noise that originates at the observation stage remains largely unexplored. Incorporating atmospheric effects into a unified system optimization of observation, communication, and semantic processing constitutes a key open research challenge.

\subsection{Data processing} 
There is a plethora of image processing and compression techniques that can be applied to the acquired images, ranging from traditional methods like JPEG to advanced \gls{ai} semantic extraction algorithms designed for specific tasks. This also includes popular object detection and localization algorithms, such as the YOLOv8 algorithm mentioned earlier and, more recently, the use of generative \gls{ai}\cite{2025FM_EO}. The selection of the specific algorithm depends on multiple factors, including the compression factor, the complexity of the algorithm, and the accuracy of the task. The compression factor indicates the reduction of the data size, which is critical for efficient data storage and transmission. The algorithm complexity affects the time elapsed from data capture until the information is ready for transmission, as well as the energy consumed during processing. More complex algorithms require additional resources but may provide better performance. Finally, task accuracy is measured by specific performance parameters that ensure the desired outcomes are met. \gls{ai} algorithms can be trained on \gls{eo} image datasets that often include images captured under varying conditions and featuring a small \gls{gsd} to achieve a high accuracy \cite{ships-google-earth_dataset}.

Regarding object detection and localization algorithms, several performance metrics are used to evaluate their effectiveness, with the supported \gls{fps} being crucial for real-time applications. Precision and recall refer to the proportion of correctly detected objects among all detected objects and among all actual objects, respectively. In addition, the \gls{map}, ranging between 0 and 1, is derived from the precision and recall values. Another metric is the \gls{iou}, which measures the overlap between the predicted boundary and the real object boundary (i.e., the ground truth). An \gls{iou} threshold is predefined (typical values being 0.5 and 0.95) to determine whether a detection is correct.

\section{System description} \label{sec:systemmodel}
 
In the considered network, the space segment is responsible for sensing (image capturing), preliminary data handling, and on-board processing, and is organized into two logically distinct but tightly coupled functional layers:  the \emph{Observation Layer}, for Earth data acquisition, and the \emph{Edge Layer}, for distributed computing, storage, and communication. These layers define functional roles rather than physically distinct satellite platforms.
Typically,  \gls{eo} satellites have limited on-board computing and storage, being optimized for sensing performance, including spatial resolution, coverage, and revisit time, rather than onboard processing and long-term storage. They transmit raw or minimally processed data (the images) to the edge layer for processing and delivery to the ground segment, where the \gls{gs} and end-users are located. However, other missions may adopt more integrated designs, with sensing and processing capabilities coexisting on the same spacecraft. 
This layered abstraction enables a general and technology-agnostic system analysis based on a logical separation between sensing and processing functionalities, while still allowing flexible cooperation between satellites.

\begin{figure}[t]
\centering
\includegraphics[width=4.5 in] {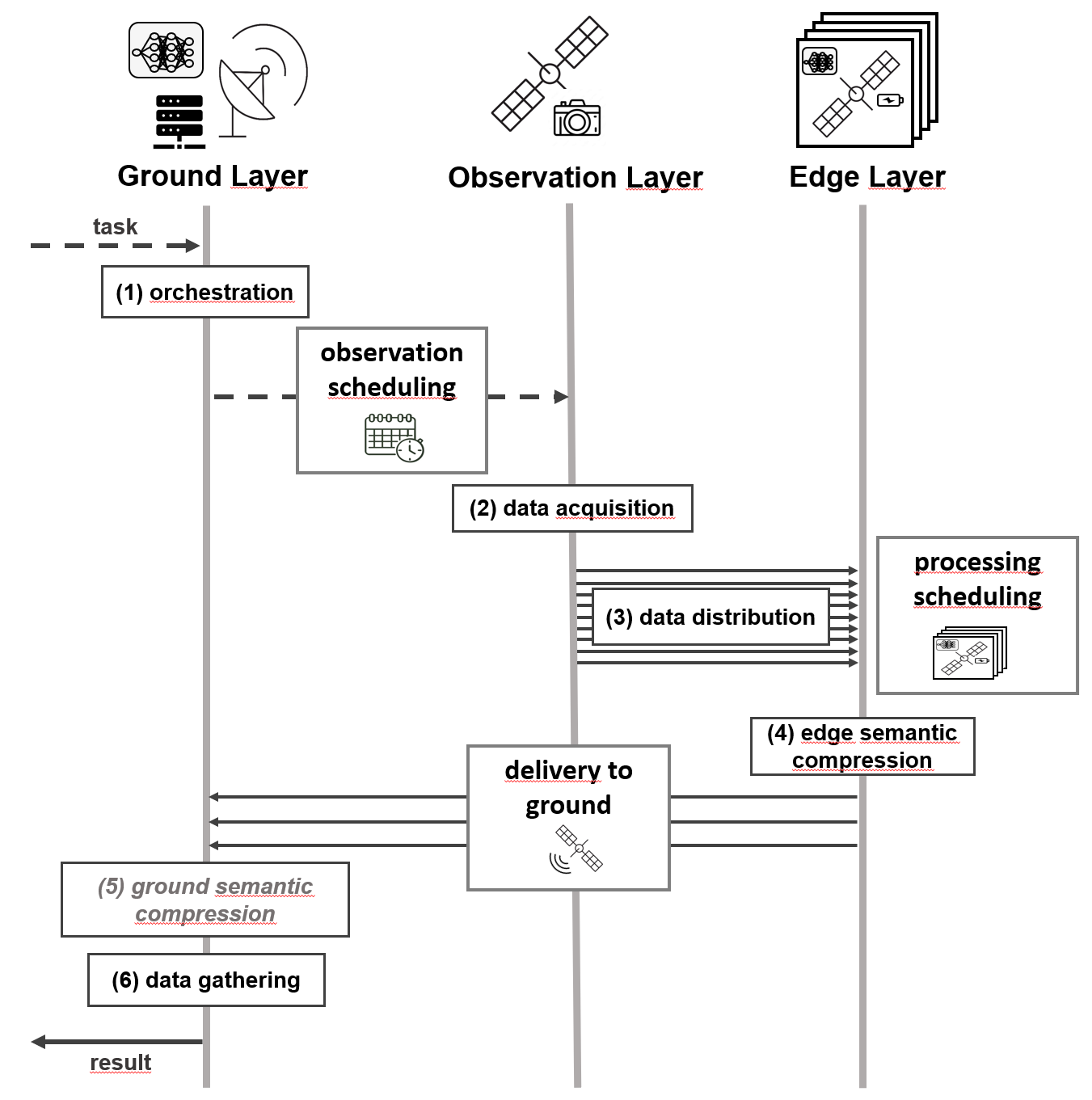}
\caption{
E2E procedure of the proposed framework: (1) Upon receiving a task, the ground station orchestrates sensing, computation, and communication resources; (2) \gls{eo} satellites acquire images according to the observation schedule; (3) the acquire data is transmitted to the edge layer; (4) based on the processing schedule, the edge layer performs semantic processing and delivers the processed data to ground; (5) any data not processed on-board is further computed at the ground layer; (6) the ground layer performs data gathering, producing the final result for the end user.} \vspace{-0.4cm} \label{fig:semanticarchitecture}
\end{figure}

The end-to-end workflow of the system (Fig.~\ref{fig:semanticarchitecture}) can be described as a sequence of interconnected functional blocks across the different layers. 
An end-user issues a query, which specifies an observation task that can be either semi-static (``Map area $X$ for $Y$ months'') or dynamic (``Track vessel with ID $X$'' or ``Map the area around volcano $Y$ in eruption''). In our scenario, the task is dynamic, involving the observation of multiple targets over a specified time horizon. 
The query is translated into a technical request to the network in the form of an orchestration of sensing, energy, computation, and communication resources, typically performed at the \gls{gs}. This orchestration is divided into observation scheduling and processing scheduling, both representing complex optimization problems~\cite{eobook}, as they must account for resource availability, network conditions, and environmental factors such as metheorological effects.
Following the observation schedule decision, which includes the attitude parameters to be applied, the \gls{eo} satellites acquire raw observational data. These measurements are affected by atmospheric noise, which impacts the quality of the acquired information, while the observation strategy aims at maximizing observation profit. The acquired data, consisting of images containing the observed targets, is then transmitted from the observation layer to the edge layer through inter-satellite links. 
{At the edge layer, semantic-aware processing is performed in parallel by cooperating edge nodes according to the processing schedule.  The images are processed and compressed using either classical algorithms, such as JPEG, or more advanced semantic-empowered processing techniques (see Table~\ref{tab:soa}), enabling task-oriented data reduction prior to transmission of the processed information to the \gls{gs}. The workload associated with semantic processing and compression is distributed across multiple edge nodes, each characterized by heterogeneous computing, energy, and connectivity capabilities. The allocation of processing tasks to edge nodes is a key system decision that must account for the available computational resources at each node, the current network conditions, and the timing constraints imposed by the end-user query.}
The processed outputs generated by the edge nodes are forwarded to the \gls{gs} through the downlink, where they are gathered. If complete processing at the edge is not feasible due to timing constraints, the remaining data is processed at the cloud-enabled \gls{gs}. 
The final result is then delivered to the end-user. For this workflow, satellites in the Observation and the Edge layer are connected to the ground segment through feeder links, in the downlink and uplink directions, while intra-space communications are enabled via intra- and cross-layer \glspl{isl}.

We next describe the models and notation for the elements of Fig.~\ref{fig:semanticarchitecture}. 

\subsection{Topology}
\noindent{\textbf{Observation Layer:}}
The \gls{aeos} satellites in the Observation Layer, indexed by $s \in \mathcal{S}$, perform data acquisition over a set of targets $\mathcal{T}$ distributed across the visible surface. The satellites are deployed at an altitude $h_o$ with an inclination $i_o$. The quality of observations is affected by atmospheric turbulence, as detailed in Section~\ref{sec:eo}. Each target $t \in \mathcal{T}$ can be observed by a satellite $s \in \mathcal{S}$ across multiple orbital passes; we denote by $O_{s,t}$ the set of orbits during which satellite $s$ can observe target $t$. As representative cases, our evaluations in Section~\ref{sec:results} consider two configurations for this layer: a single \gls{aeos} and a four-satellite Walker Star constellation, with one satellite per orbital plane.  

\noindent{\textbf{Edge Layer:}}
The Edge Layer consists of a \gls{leo} satellite constellation of $N$ cooperative satellites, indexed by $e \in \mathcal{E}$. These satellites process raw or partially processed data close to the source, supporting timely and efficient delivery of results to the \gls{gs}.
Without loss of generality, we consider a single orbital plane deployed at an altitude $h_e$ with an inclination $i_e$, noting that the extension to multiple planes follows directly. The slant range between these satellites is defined as $d_\text{ISL} = 2(R_E+h_e)\sin{(\pi/N)}$, where $R_E$ denotes the Earth's radius. Each satellite in this layer is equipped with computational resources to perform onboard processing, as detailed later, and communicates with both the Observation and the Ground Layer.

\noindent{\textbf{Ground Layer:}}
The ground segment comprises a set of \glspl{gs}, denoted by $g \in \mathcal{G}$, which connect to end-users and the space segment via feeder links, in the downlink and uplink directions. Without loss of generality, we assume a single \gls{gs} receives user requests, which are translated into a resource orchestration plan that includes the observation scheduling and processing allocation, and all \glspl{gs} can potentially receive processed images from the Edge Layer. Ground stations are equipped with substantial cloud computing resources to handle non-time-critical tasks and to augment edge capacity. The downlink feeder links connecting the \glspl{gs} to the Edge Layer are typically the main communication bottleneck. In the evaluations of Section\ref{sec:results}, the $26$ \glspl{gs} are located at ground positions accordingly to the KSAT ground station service~\cite{KSATGroundNetwork}. 

\subsection{Acquisition model}
For each orbit $o \in O_{s,t}$ of the Observation Layer satellites, $\text{VTW}_{s,t,o}$ represents the \gls{vtw} for satellite $s$ over target $t$, defined as the interval $[\text{sw}_{s,t,o}, \text{ew}_{s,t,o}]$, with $\text{sw}_{s,t,o}$ and $\text{ew}_{s,t,o}$ representing its start and end times. 
Exploiting the enhanced maneuverability of \glspl{aeos}, each $\text{VTW}_{s,t,o}$ is discretized into a set of \glspl{otw} with a fixed time step of $\text{prc}$ seconds. Accordingly, $\text{OTW}_{s,t,o,w}$ denotes the \gls{otw} for satellite $s$ and target $t$ in orbit $o$, and $w \in W_{s,t,o}$ refers to the indexes of \glspl{otw} contained within such $\text{VTW}_{s,t,o}$. 
We assume that the actual observation duration is significantly shorter than the required attitude maneuvering and onboard processing times, which is typically the case; hence, the operation is constrained by these tasks rather than by acquisition time.
Accordingly, each $\text{OTW}_{s,t,o,w}$ is characterized solely by its observation timestamp $\tau_{s,t,o,w}$ and an associated profit $\sigma_{s,t,o,w}$. For the sake of simplicity, this profit is defined in terms of image quality, specifically the \gls{gsd}, although the formulation can be extended to incorporate additional \glspl{kpi} such as image freshness.~\cite{mercadomartinez2025schedulingagileearthobservation}. 
  
Observing $\text{OTW}_{s,t,o,w}$ requires a given attitude defined by its roll ($\theta_{s,t,o,w}$), pitch ($\phi_{s,t,o,w}$), and yaw ($\psi_{s,t,o,w}$) angles. Each satellite's maneuvering capabilities  are restricted by  its maximum roll, pitch, and yaw angles ($\theta_{\text{max}}$, $\phi_{\text{max}}$, and $\psi_{\text{max}}$, respectively), as well as by the slewing velocity of the sensor. The attitude transition time function is typically modeled as a piecewise linear function. Specifically, the attitude transition time between two consecutive observations is computed as~\cite{LIU201741}:
\begin{equation}
    \Delta \tau_{t,w-t',w'} = 
    \begin{cases}
        11.66, & \alpha_{t,w-t',w'} \leq 10^\circ \\
        5 + \alpha_{t,w-t',w'} / 1.5, & 10^\circ < \alpha_{t,w-t',w'} \leq 30^\circ \\
        10 + \alpha_{t,w-t',w'} / 2, & 30^\circ < \alpha_{t,w-t',w'} \leq 60^\circ \\
        16 + \alpha_{t,w-t',w'} / 2.5, & 60^\circ < \alpha_{t,w-t',w'} \leq 90^\circ \\
        22 + \alpha_{t,w-t',w'} / 3, & \alpha_{t,w-t',w'} > 90^\circ \\
    \end{cases}
    ,
    \label{eq:attitude_transition_time}
\end{equation}
\begin{equation}
    \alpha_{t,w-t',w'} = |\theta_{s,t,o,w} - \theta_{s,t',o,w'}| + |\phi_{s,t,o,w} - \phi_{s,t',o,w'}| + |\psi_{s,t,o,w} - \psi_{s,t',o,w'}|
    \label{eq:angle_displacement}
    ,
\end{equation}
where $\alpha_{t,w-t',w'}$ is the total attitude transition angle between $\text{OTW}_{s,t,o,w}$ and $\text{OTW}_{s,t',o,w'}$. The power required for attitude maneuver transitions is denoted by $P_{\text{man}}$. While existing literature on the \gls{aeossp} frequently discusses the energy consumption rate (energy per unit of time), we explicitly formulate this in terms of power to maintain consistency throughout this work. For each satellite $s$, the total maneuvering energy is limited by $E_{\text{max}}$, which defines the maximum allowable energy budget for attitude transitions. 

The sequence of selected $\text{OTW}_{s,t,o,w}$ by the \gls{aeossp} is denoted by $\mathrm{OTW}^\star$, and it defines the sequence of times $\tau_{s_k,t_k,o_k,w_k}$ at which the Observation Layer will deliver a new image to be processed by the Edge Layer:

\begin{equation}
\label{eq: yfirst}
\begin{aligned}
\mathrm{OTW}^\star = &
\big( \mathrm{OTW}_{s_1,t_1,o_1,w_1}, \mathrm{OTW}_{s_2,t_2,o_2,w_2}, \dots, \mathrm{OTW}_{s_K,t_K,o_K,w_K} \big), \\
&\tau_{s_1,t_1,o_1,w_1} \le \tau_{s_2,t_2,o_2,w_2} \le \dots \le \tau_{s_K,t_K,o_K,w_K}, \\
&\;\;y_{s_i,t_i,o_i,w_i} = 1, \ \forall i = 1,\dots,K,
\end{aligned}
\end{equation}

\noindent where $y_{s_i,t_i,o_i,w_i}$ is the binary decision variable that denote if $\mathrm{OTW}_{s_i,t_i,o_i,w_i}$ is scheduled or not. 
Each element in $\mathrm{OTW}^\star$ corresponds to an observation, i.e., a frame composed of $N_\text{img}$ images of size $D_\text{img}$ with a total size $D_s$ to be processed by the Edge Layer. The images containing the targets have a resolution 
$w_\text{img} \times h_\text{img}\,\text{bits}$, where $w_\text{img}$ and $h_\text{img}$ are the width and height in pixels.

As described in Section~\ref{sec:eo}, atmospheric turbulence significantly affects the quality of this acquired data and is quantified by the $C_n^2$ value during the \gls{otw}, which in our evaluations is sampled from the distribution in Fig.~\ref{fig:cn2_cdf}. 
To ensure high-quality observations, we establish a threshold, {$C_{n, \text{max}}^2(0)$}, measured at ground-level, which represents the maximum allowable turbulence for a high-quality frame. Any observation performed under conditions exceeding this threshold is discarded.

{\subsection{Computation model}
\label{subsec:comp_energy_model}
We next introduce the computational and energy consumption models used to characterize the processing capabilities of the Edge and Ground Layers nodes, providing an abstract representation of execution time and energy consumption. 
The space and ground processing nodes, indexed by $p \in \mathcal{P}=\set{E}\cup\set{G}$
, are equipped with a multi-core parallel processing architecture, characterized by the number of available processing cores $N_\text{cores}^{(p)}$, the maximum operating frequency $f_\text{max}^{(p)}$, and the power consumption $P_\text{max}^{(p)}$ when operating at maximum frequency. In the evaluations of Section~\ref{sec:results}, both \gls{cpu} and \gls{gpu} architectures are considered, resulting in heterogeneous parameter settings. 

When processing node $p$ operates at a clock frequency $f_p \leq f_\text{max}^{(p)}$, its processing power consumption is modeled as
\begin{equation}
P\left(f_p\right)=P_\text{max}^{(p)}\left(\frac{f_p}{f_\text{max}^{(p)}}\right)^3.
\end{equation}
Let $W$ denote the \emph{work} of the algorithm of interest, defined as the number of \glspl{flop} required to process a single unit of data (i.e., an image). The value of $W$ is assumed to be known and depends on the algorithmic characteristics, such as the stopping criteria or the architecture of a neural network in a machine learning model.
To account for non-ideal execution, we introduce a \gls{rv} $C^{(p)}$ 
 that captures the computational inefficiency at processing node $p$ when executing a computing task. This factor models the overhead due to imperfect task allocation, memory access, and data management, and depends on both the node architecture and the algorithm.
Finally, let $N_\text{FLOPs}^{(p)}$ denote the number of \glspl{flop} that each processing core at node $p$ can perform per clock cycle.

According to the \gls{bsp} model~\cite{Amaris15}, the execution time of an algorithm that is executed in parallel in processing node $p$ operating at frequency $f_p$ can be modeled as
\begin{IEEEeqnarray}{rCl}
T_\text{proc}^{(p)}\left(f_p\right)&=&T_{\text{work}}^{(p)}\left(f_p;C^{(p)}\right)+T^{(p)}_\text{sync},\IEEEeqnarraynumspace
\label{eq:t_proc}
\end{IEEEeqnarray}
where \gls{rv} $T_\text{sync}^{(p)}$ 
is the delay due to communication and synchronization among processing cores in the architecture, and the \gls{rv} of the time needed to execute the work of the algorithm is
\begin{equation}
T_{\text{work}}^{(p)}\left(f_p;C^{(p)}\right)=\frac{C^{(p)}W}{N_\text{cores}^{(p)}N_\text{FLOPs}^{(p)}f_p}.
\end{equation}
Consequently, the mean execution time $\mu_T^{(p)}$ is modeled as
\begin{IEEEeqnarray}{rCl}
\mu_T^{(p)}\left(f_p\right)&=&\frac{\mu_C^{(p)}W}{N_\text{cores}^{(p)}\,N_\text{FLOPs}^{(p)}\,f_p}+\mu^{(p)}_\text{sync},\IEEEeqnarraynumspace
\label{eq:mean_t_proc}
\end{IEEEeqnarray}
\noindent where $\mu_C^{(p)}$ and $\mu_\text{sync}^{(p)}$ are the mean values of $C^{(p)}$ and $T_\text{sync}^{(p)}$, respectively. $T_\text{proc}^{(p)}(f_p)$ is experimentally characterized in Section~\ref{sec:results} for representative computing platforms.

If the processing power for a given clock frequency $P(f_p)$ is constant, the mean energy consumption for processing an image at node $p$ using an algorithm requiring $W$ \glspl{flop} can be calculated as
\begin{IEEEeqnarray}{rl}
    E_{\text{proc}}^{(p)}\left(f_p; W\right) &= P\left(f_p\right)\cdot \mu_T^{(p)}\left(f_p\right)=\frac{P_\text{max}^{(p)} f_p^3\, \mu_T^{(p)}\left(f_p\right)}{\left(f_\text{max}^{(p)}\right)^3}\! \IEEEeqnarraynumspace\IEEEnonumber\\
     &=\frac{P_\text{max}^{(p)}}{\left(f_\text{max}^{(p)}\right)^3}\!\left(\frac{f_p^2 \mu_C^{(p)}W}{N_\text{cores}^{(p)}N_\text{FLOPs}^{(p)}}+f_p^3\mu^{(p)}_\text{sync}\right)\!.\IEEEeqnarraynumspace
\end{IEEEeqnarray}

\subsection{Communication model}
\label{sec:comms_model}
The satellites are interconnected both with the ground segment and with each other through the feeder link and the \glspl{isl}, respectively. 
\Gls{fso} technology is considered the primary \gls{isl} solution, as it has become the \emph{de facto} standard for high-data-rate \gls{leo} inter-satellite communications. Due to the controlled relative positioning and stable distances among satellites within the constellation, pointing errors can be effectively minimized. Consequently, \glspl{isl} are assumed to be fixed during network operation, with a constant data rate $R_{\mathrm{ISL}}$ and consuming a fixed transmission power $P_{\mathrm{ISL}}$. In contrast, satellite-to-ground \gls{rf} feeder links are more robust to atmospheric impairments and less sensitive to pointing errors than \gls{fso}, which makes them suitable for \gls{dl} transmissions where channel conditions vary due to satellite mobility and atmospheric effects. Hence, the feeder links are assumed to rely on conventional \gls{rf} technology, characterized by bandwidth $B$, transmission power $P_{\mathrm{DL}}$, and total antenna gain $G_{\mathrm{DL}}$.

The system operation is divided into time slots of equal duration $T_\text{slot}$, which are indexed by $k\in\{0,1,2,\dotsc\}$.
Let $N_\text{img}(k)$ be the number of images captured at time slot $k$ by the Observation layer, with a total data size $D_s$. 
  Moreover, for notation simplicity,{let $g(k)$ denote the ground station defined for transmitting the images captured at slot $k$} in the downlink  and $\ell_g(k)=(e,g(k))$ be the (edge) satellite-to-ground link established to download the data captured at time slot $k$, such that $e\in\set{E}$ is the satellite that communicates with ground station $g(k)$. We define the time-varying data rate for $\ell_g(k)$ from its \gls{snr} on an interference-free \gls{awgn} channel with free-space path loss and with noise power $\sigma$ as
\begin{equation}
    \gamma_k = G_\text{DL}\,P_\text{DL}\left(\frac{\mathrm{c}}{4\pi\, d_{e,g}(k) \,f_c\,\sigma}\right)^2,
\end{equation}
where $f_c$ is the carrier frequency, and $d_{e,g}(k)$ is the distance between satellite $e$ and \gls{gs} $g(k)$.
Then, the communication rate is selected using the set of achievable spectral efficiencies defined in the DVB-S2X system~\cite{dvb_s2}, which constitute the set $\mathcal{R}_\text{DVB}=\left\{r\right\}$ in b/s/Hz. Next, let $\gamma_\text{min}(r)\geq 2^{r}-1$ be the minimum required \gls{snr} to achieve a block error rate $<10^{-5}$ with spectral efficiency $r$. The modulation and coding scheme for downlink communication is selected to achieve the data rate
\begin{equation}
    R_k=B\max \left\{r\in\set{R}_\text{DVB-S2}:\gamma_{k}\geq \gamma_\text{min}(r)\right\}.
\end{equation}

\emph{Scattering} refers to the transmission of \emph{uncompressed} data from a source \gls{eo} satellite $s \in \mathcal{S}$ to a processing node $p \in \mathcal{P}=\mathcal{E} \cup \mathcal{G}$, i.e., either to an edge-layer satellite or directly to a ground station for processing.
{\emph{Gathering}, on the other hand,  is the operation of collecting the \emph{compressed} data from all the servers at the ground station. In the following, the apex $(u)$ will be the label associated to the scatter phase where the data is \emph{uncompressed}, while the apex $(c)$ the label associated to the gather phase where the data has been \emph{compressed}.} 

Immediately after capturing a frame of $N_\text{img}$ images at satellite $s$ with a total data size $D_s$, these are scattered to the edge satellites for processing, leading to a distributed computing problem. The goal is to deliver the data in a timely manner while reducing the \gls{e2e} energy consumption, which involves the joint optimization of communications and computing.  To maintain the stability of the system, the scatter phase for an observation performed at time slot $k\in\{0,1,\dotsc\}$ is scheduled in the same time slot $k$. Then, the processing of the $N_\text{img}$ captured images is scheduled in time slot $k+1$, and the gather phase is scheduled in time slot $k+2$. Note that this allows for implementing pipelining, where the scatter phase of the images captured at time slot $k$ occurs in parallel with the processing and gathering phases for images captured at time slots $k-1$ and $k-2$, respectively.} 

{The overall path from satellite $s$ to ground station $g$ is time-varying. During scattering and gathering, data may need to traverse multiple \glspl{isl} before reaching a satellite that is currently connected to the ground segment. While the \glspl{isl} are assumed to be stable and fixed during network operation, the ground-to-satellite feeder link may change over time. In particular, both the satellite $e\in\set{E}$ connected to the \gls{gs} via the feeder link $e_g(k)$ and the corresponding transmission rate $R_k$ may vary with time index $k$.
Hence, time dependence is explicitly captured in the notation through the time-slot index $k$. Accordingly, the path for sending uncompressed data from $s$  to $p$ and compressed data from $p$ to $g$ in time slot $k$ is denoted as
\begin{equation}
\mathcal{L}_{s,p,g}(k)=\mathcal{L}^{(u)}_{s,p}(k)\bigcup\mathcal{L}^{(c)}_{p,g}(k),\quad p\in\mathcal{P}.
\end{equation}
where $\mathcal{L}^{(u)}_{s,p}(k)$ is the set of links where  uncompressed bits are routed to satellite $p$ (\emph{scatter phase}) and  $\mathcal{L}^{(c)}_{p,g}(k)$  is the set of links over which the compressed  bits are routed  from satellite $p$ to the ground station (\emph{gather phase}) in time slot $k$.  
}

During the scatter phase, the observation satellite transmits data packets -- each containing entire images of average size $D_\text{img}$ or segments of them -- to the edge satellites. Let $D\in\left(0, D_\text{img}\right]$~bits be the size of a data packet to be transmitted. 
Since the intra-plane \gls{isl} data rate, $R_{\text{ISL}}$, is fixed, the mean time needed to deliver and image from the source satellite $s$ to a processing satellite $p\in\set{E}$ is the sum of the transmission time and the propagation time, i.e., 
\begin{equation}
T_\text{comm}^{(u)}{(s,p, D)}=\sum_{\ell\in\mathcal{L}_{s,p}^{(u)}(k)} \left(\frac{D}{R_\text{ISL}}+\frac{d_\text{ISL}}{c}\right)
\end{equation}

Conversely, the time needed to deliver the semantic information extracted from an image in satellite $p$ to the \gls{gs} $g$ in time slot $k$ is
\begin{IEEEeqnarray}{rCl}
T_\text{comm}^{(c)}{(p, g,D)}&=&
T_\text{comm}^{(c)}(p,e,D)+T_\text{comm}^{(c)}(e,g,D)\\
&=&
\sum_{\ell\in\mathcal{L}_{p,e}^{(c)}(k)}\left(\frac{D}{\rho R_\text{ISL}}+\frac{d_\text{ISL}}{c}\right)+\frac{D}{\rho R_k}+\frac{d_{e,g}(k)}{c},
\label{eq:comm_latency}
\end{IEEEeqnarray}
where $\rho\geq 1$ is the summarization capacity (i.e., compression ratio) of the compression algorithm.

Next, we define the energy consumption models for communication. Since the \glspl{isl} operate at a fixed transmission power and we assume an ideal power amplifier efficiency, the energy required to transmit $D$~bits of data over the $\ell$-th \gls{fso} \gls{isl} is given by the product of the effective transmit power and the transmission duration
\begin{equation}
    E_{\text{ISL},\ell}^{\text{tx}}(D) = \frac{P_{\text{ISL},\ell} D}{R_{\text{ISL}}}.
\end{equation}
Similarly, the energy required to transmit $D$~bits of data over the feeder link during time slot $k$ is
\begin{equation}
    E_\text{DL}^{\text{tx}}(D,k) = \frac{P_{\text{DL}} D}{R_k}.
\end{equation}

\section{Optimization} \label{sec:optimization}
Fulfilling an \gls{eo} task in a constrained environment involves a delicate balance between accuracy, delay, and energy consumption. The conflicting objectives  must all  be considered in the resource allocation optimization, which can be conceptually formulated as follows: \emph{Identify the best set of \gls{eo} and edge satellites for image capturing, semantic information extraction and encoding, and routing to the ground, with the aim of minimizing energy consumption while meeting accuracy and timing constraints}. 
Energy consumption is influenced mainly by the data acquisition, on-board algorithm execution, and data transmission. Image capturing is mainly impacted by the amount and frequency of the targets to be observed, which can exceed the Observation Layer capacity. On the Edge Layer, semantic extraction introduces a fundamental trade-off: it increases onboard processing energy, yet significantly reduces transmission energy by shrinking the data volume, since semantically processed information is much smaller than raw imagery.
Additionally, the system optimization must take into account the hard constraint that each frame must be captured, processed, and completed before the subsequent frame arrives. If this is not achieved, images may need to be discarded or not captured to avoid congestion in the system. If only a portion of the data can be processed, the remaining data will be transmitted in its raw form to the \gls{gs}, thereby increasing the load on the communication network and prolonging transmission time.

These considerations form the basis for formulating the two optimization problems underlying observation scheduling ($\mathrm{P}_{\text{obs}}$) and processing scheduling ($\mathrm{P}_{\text{sch}}$), as discussed next. 

\subsection{Observation scheduling and attitude control for observation profit maximization}

The main difficulty when addressing the \gls{aeossp}, denoted $\mathrm{P}_{\text{obs}}$, is that the volume of requests exceeds the observation capacity of the \gls{aeos} and therefore only a selection of targets can be observed in each time slot. In this case, the aim is to maximize the collected observation profit while meeting all time, energy, and storage constraints. Data from the Observation Layer is dynamically processed as it is acquired, and therefore storage constraints are neglected in this problem.

As introduced in \eqref{eq: yfirst}, for each observation time window $\text{OTW}_{s,t,o,w}$, let
$y_{s,t,o,w} \in \{0,1\}$ denote whether the window is scheduled ($y_{s,t,o,w}=1$) or not ($y_{s,t,o,w}=0$). 
Furthermore, let $z_{t,w,t',w'} \in \{0,1\}$ 
indicate whether $\text{OTW}_{s,t,o,w}$ is scheduled immediately before $\text{OTW}_{s,t',o,w'}$.
The goal is to optimize the profit associated with observation $\text{OTW}_{s,t,o,w}$, which we define in terms of image quality. Without loss of generality, we assume that all targets have the same priority and define the profit in terms of the \gls{gsd} during the observation as
\begin{equation}    \sigma_{s,t,o,w} = \frac{\text{GSD}_{\text{nadir}}}{\text{GSD}_{s,t,o,w}},
\label{eq:observation_profit}
\end{equation}
where $\text{GSD}_{\text{nadir}}$ and $\text{GSD}_{s,t,o,w}$ are the \gls{gsd} at nadir and during $\text{OTW}_{s,t,o,w}$, respectively.

The observation scheduling problem is then formulated as 

\begin{IEEEeqnarray*}{rrCll}
    \label{eq:aeossp}
    \mathrm{P}_{\text{obs}}: &\maximize_{y_{s,t,o,w},\, z_{t,w,t',w'} \in \{0,1\}} & \quad & \IEEEeqnarraymulticol{2}{l}{
        \sum_{s \in S} \sum_{t \in T} \sum_{o \in O_{s,t}} \sum_{w \in W_{s,t,o}} \sigma_{s,t,o,w} \cdot y_{s,t,o,w}
    }
    \IEEEyesnumber\label{eq:objective_function}\\
    &\text{subject to} & \quad & \text{sw}_{s,t,o} \leq \tau_{s,t,o,w} \leq \text{ew}_{s,t,o}
    \IEEEyessubnumber\label{eq:window_constraint}\\
    &&& {z_{t,w,t',w'} \cdot \left(\tau_{s,t,o,w} + \Delta \tau_{t,w-t',w'}\right)} \leq \tau_{s,t',o,w'}
    \IEEEyessubnumber\label{eq:attitude_constraint}\\
    &&& \sum_{t, t' \in T} \sum_{w, w'\in W_{s,t,o}}z_{t,w,t',w'} \cdot \left( P_{\text{man}} \cdot \Delta \tau_{t,w-t',w'} \right) \leq E_{\text{max}}
    \IEEEyessubnumber\label{eq:energy_constraint}\\
    &&& \sum_{s \in S} \sum_{o \in O_{s,t}} \sum_{w \in W_{s,t,o}}{y_{s,t,o,w}} \leq 1
    \IEEEyessubnumber\label{eq:binary_variable_constraint}\\
\end{IEEEeqnarray*}
where the optimization objective function (\ref{eq:objective_function}) is designed to maximize the sum of collected profits; constraint  (\ref{eq:window_constraint}) ensures that the observation of target $t$ occurs within one of its \gls{vtw}; constraint (\ref{eq:attitude_constraint}) models the attitude transition requirement,  ensuring that $\text{OTW}_{s,t',o,w'}$ can only be scheduled after $\text{OTW}_{s,t,o,w}$ if the attitude maneuver is feasible; constraint (\ref{eq:energy_constraint})  guarantees that the total energy consumption does not exceed the maximum allowed energy and  constraint (\ref{eq:binary_variable_constraint}) guarantees that each target can be observed at most once in the schedule.

In $\mathrm{P}_{\text{obs}}$, the integer optimization variables take values in $\{0,1\}$, and both the objective function and the constraints are linear; therefore, this is a \emph{binary integer linear programming} (0–1 ILP) problem. Such formulations, which are commonly used to model resource-constrained scheduling tasks, do not admit closed-form solutions and can exhibit high computational complexity, especially at scale. Therefore, sub-optimal or heuristic approaches are typically adopted in practice, as exhaustive search becomes computationally prohibitive for large-scale instances. In this work, we solve $\mathrm{P}_{\text{obs}}$  using the SCIP solver~\cite{Achterberg2009, 10.1007/978-3-540-68155-7_4}, a state-of-the-art framework for mixed-integer optimization.

\subsection{Edge computing for energy minimization}
The scheduling of edge computing and downlink tasks is organized according to the available downlink times at the ground segment and the observation scheduling for image acquisition at the \gls{eo} satellites. 
The pipeline model establishes that $N_\text{img}$ captured and scattered in time slot $k$ are processed in $k+1$ and gathered in time slot $k+2$. This imposes a constraint for the optimization that the scattering and processing of the images in a frame captured at time slot $k$ must complete in time to reach the downlink satellite before the end of time slot $k+2$. Otherwise, downlink is considered a failure and the system must proceed to capture, process, and deliver the frames captured in the upcoming time slots. 

The optimization variables are the $P$-dimensional vector of the processing frequencies $\mathbf{f}=[f_1,f_2,\dots,f_P]$ of each satellite and \gls{gs} in $\mathcal{P}$ and the $SP$-dimensional  vector of the load distribution $\mathbf{x}=[x_{1,1},x_{1,2},\dots,x_{S,P}]$. The variable $x_{s,p}$ indicates the fraction of the computation load generated in $s\in\mathcal{S}$ that is processed in $p\in \mathcal{P}$. Accordingly, a value of $x_{s,p}\ne 0$ means that  $x_{s,p}D_s$ uncompressed  bits are scattered from the source satellite $s$ to the processing node $p$ and then $x_{s,p}D_s/\rho$ compressed bits are gathered from satellite $p$ to ground station $g$. 
Both the computing load and data scale linearly with the value of $x_{s,p}$, so that if source $s$ generates the load $C_s$ and data $D_s$, the load processed at the satellite $p$ is $x_{s,p}C_s$  addressing  $x_{s,p}D_s$ data.

The mean energy consumption for a batch of images captured at time slot $k$ is modeled as the sum of the transmission energy and the processing energy, i.e.,
    \begin{IEEEeqnarray*}{rCl}
        E_k\left(\mathbf{x},\mathbf{f};W,D_s,\rho\right)&=&E_k^\text{(u)}\left(s,g,\mathbf{x},D_s\right) 
        +      E_{k}^\text{(c)}\left(s,g,\mathbf{x},D_s,\rho\right)+\sum_{p\in\mathcal{P}} \frac{D_sx_{s,p}}{D_\text{img}}E_{\text{proc}}^{(p)}\left(f_p;W\right).\IEEEyesnumber
\end{IEEEeqnarray*}

Given the workload allocation vector $\mathbf{x}$, the total energy consumed during the scattering phase, which occurs during the observation time slot $k$, corresponds to the energy required to transmit the uncompressed data to the processing nodes through the routing paths $\left\{\set{L}_{s,p}^{(u)}(k)\right\}$, including the downlink to the \gls{gs} $g(k)$ as
 \begin{equation}
     E^\text{(u)}_k\left(s,g,\mathbf{x},D_s\right)=\sum_{p\in\mathcal{E}}\left(\sum_{\ell \in \mathcal{L}^{(u)}_{s,p}(k)}E_{\text{ISL},\ell}^\text{tx}\left(x_{s,p}D_s\right)\right)+\sum_{\ell\in\set{L}^{(u)}_{s,e}(k)}E_{\text{ISL},\ell}^\text{tx}\left(x_{s,g(k)}D_s\right)+ E_{\text{DL}}^\text{tx}\left(x_{s,g(k)} D_s, k\right),
     \end{equation}   
 Similarly, the energy consumed during the gathering phase corresponds to the energy required to transmit the compressed data to the ground station.\begin{equation}
 E^\text{(c)}_{k}\left(s,g,\mathbf{x},D_s,\rho\right)=\sum_{p\in\mathcal{E}}\left(\sum_{\ell \in \mathcal{L}^{(c)}_{p,g}(k)}E_{\text{ISL},\ell}^\text{tx}\left(x_{s,p}\frac{D_s}{\rho}\right)+E_{\text{DL}}^\text{tx}\left(x_{s,p} \frac{D_s}{\rho}\right)\right)
 \end{equation}
Accordingly, the scheduling optimization  Problem $\mathrm{P}_{\text{sch}}$, which determines the optimal task allocation vector $\mathbf{x}$ and processor frequency vector $\mathbf{f}$, is formulated based on the \gls{bsp} model as
    
 \begin{IEEEeqnarray*}{rrCll}
    \mathrm{P_{\text{sch}}}:\,&\minimize_{\mathbf{x},\mathbf{f}}& \quad &\IEEEeqnarraymulticol{2}{l}{E_k\left(\mathbf{x},\mathbf{f};W,D_s,\rho\right)}
    \IEEEyesnumber\label{eq:optimize_distr_single}\\
    &\text{subject to} &\quad & \frac{D_sx_{s,p}\mu_T^{(p)}(f_p)}{D_\text{img}}\leq  T_\text{slot} & \quad \forall p\in{\set{P}}\IEEEyessubnumber\label{c:proc}\\
    &&& \IEEEeqnarraymulticol{2}{l}{D_s\left(x_{s,g}+\frac{1}{\rho}\sum_{p\in{\set{E}}} x_{s,p}\right) \leq (T_\text{slot}-T_\text{delay}(k))R_{k}}\IEEEyessubnumber\label{c:dl}\\
    &&&\sum_{p\in\set{P}}Dx_{s,p}\left(\mathcal{I}\left(\ell,\mathcal{L}_{s,p}^{(u)}(k)\right)+\frac{1}{\rho}\mathcal{I}\left(\ell,\mathcal{L}_{p,g}^{(c)}(k)\right)\right)
   \leq T_\text{slot}R_{\text{ISL},\ell} & \quad  \ell\in\mathcal{L}_{s,g,p}(k) \IEEEyessubnumber*\label{c:isl}\\
    &&&  f_p\leq f_p^\text{max}, &\quad \forall p\in{\set{P}}\label{c:procf}\\
    &&& 0\leq x_{s,p}\leq 1, &\quad \forall p\in{\set{P}}\label{c:scatt}\\
&&&x_{s,g}+\sum_{p\in{\set{E}}}x_{s,p} =  1. \label{c:cons}
    \end{IEEEeqnarray*}
{Constraint~(\ref{c:proc}) represents the processing latency constraint, ensuring the computation assigned to processing node $p$ in $\in\set{P}$ can be completed, on average, within one time slot. 
Specifically, the mean processing time required to handle the fraction $x_{s,p}$ of the workload generated at source $s$ must not exceed the slot duration $T_{\text{slot}}$.
Constraint~(\ref{c:dl}) 
ensures that the total data sent to the ground station during time slot $k+2$, including both compressed and uncompressed portions, remains within the available downlink capacity, where 
\begin{equation}
T_\text{delay}(k)=\max_{p\in\set{E}} \left(T_\text{comm}^{(c)}{(p, e,D_s\,x_{s,p})}+\max\left\{T_\text{proc}^{(p)}\left(f_p;D_s\,x_{s,p}/D_\text{img}\right)-T_\text{slot},0\right\}\right)
\end{equation}
is the maximum delay experienced by the data from an observation at slot $k$ to begin the downlink in the scheduled time slot $k+2$.
Constraint~(\ref{c:isl}) enforces the inter-satellite link (ISL) capacity limit. For each ISL $\ell$ involved in the routing path $\mathcal{L}_{s,g,p}(k)$ during time slot $k$, the total traffic — either  compressed or not — must not exceed the available ISL transmission capacity. The indicator function $\mathcal{I}(\ell,\mathcal{L})$ equals 1 if the routing path $\mathcal{L}$ contains link $\ell$ and 0 otherwise. It is therefore used to account only for the portion of traffic that traverses link $\ell$.
Constraint~(\ref{c:procf}) imposes bounds on the processor frequencies, ensuring that each processing node operates within its feasible frequency range.
Constraint~(\ref{c:scatt}) enforces that the workload allocation variables are fractional and bounded between zero and one.
Finally, constraint~(\ref{c:cons}) guarantees workload conservation, i.e., the total computation load generated at source $s$ is fully allocated across the ground station and the edge satellites.}

Problem $\mathrm{P_{\text{sch}}}$ is non-convex: a local-optimal solution can be obtained through partial dual decomposition, block coordinate descent, or successive convex approximation. However, we observed that the processor frequency $f_p^*$ for all $p\in\set{P}$ that minimizes the energy consumption under stability constraints~\eqref{c:proc} and~\eqref{c:procf} is 
\begin{equation}
    f_p^*=
    \frac{\mu_C^{(p)}W}{\displaystyle N_\text{cores}^{(p)}N_\text{FLOPs}^{(p)}\, \left(\frac{T_\text{slot}D_\text{img}}{D\,x_{s,p}}-\mu_\text{Tsync}^{(p)}\right)}
\end{equation}
which makes $\mu_T^{(p)}\left(f_p^*\right)=T_\text{slot}$ for all $p\in{\set{P}}$.
Thus, for a given $x_{s,p}$, the optimal processing energy is
\begin{equation}
     \frac{Dx_{s,g}}{D_\text{img}}E_{\text{proc}}^{(p)}\left(f_p^*;W\right)
     =\frac{P^{(p)}_\text{max}T_\text{slot}\left(\mu_C^{(p)}W\right)^3}{\left(\displaystyle N_\text{cores}^{(p)}N_\text{FLOPs}^{(p)}\, \left(\frac{T_\text{slot}D_\text{img}}{D\,x_{s,p}}-\mu_\text{Tsync}^{(p)}\right)f_\text{max}^{(p)}\right)^3}
     \label{eq:opt_eproc}
\end{equation}
By substituting~\eqref{eq:opt_eproc} into~\eqref{eq:optimize_distr_single} and
replacing the constraints~\eqref{c:proc} and \eqref{c:procf} with
\begin{equation}
    \frac{D_sx_{s,p}}{D_\text{img}}\left(\frac{\mu_C^{(p)}W}{N_\text{cores}^{(p)}N_\text{FLOPs}^{(p)}\, f^{(p)}_\text{max}} + \mu_\text{Tsync}^{(p)}\right)\leq  T_\text{slot}, \quad \forall p\in{\set{P}}, 
\end{equation}
 problem $\mathrm{P_{\text{sch}}}$ becomes convex and can be readily solved using standard optimization libraries.

\section{Performance evaluation} \label{sec:results}

We present exemplary results for the two optimization problems in a system comprising a satellite constellation with $23$ satellites that provide edge computing resources to one \gls{eo} satellite. Parameters from the satellite WorldView-3 \cite{worldview-13} are used for the image acquisition model and orbital parameters. Without loss of generality, we assume that the observation satellites also serve as edge node, i.e., they perform dual functions, so that $h_e = h_o = 617$  km and $i_e = i_o = 98.6^\circ$. The task of interest is vessel detection within a defined area of interest. The Observation Layer captures images of regions where vessels are present, which are then semantically processed using YOLOv8 to perform vessel localization and detection.

Attitude maneuvers are constrained by slew limits $\theta_{\max} = 45^\circ$, $\phi_{\max} = 45^\circ$, and $\psi_{\max} = 90^\circ$, and by a total energy budget of $E_{\text{max}}=1000$ J. The power consumption for attitude transitions is set to \mbox{$P_{\text{man}}=2$ W}. Additionally, aligning with WorldView-3 specifications, we assume \mbox{$\text{GSD}_{\text{nadir}}=0.31\,\text{m/pixel}$}.
We consider scenarios both with and without the effect of atmospheric turbulence. When turbulence is accounted for, we perform a Monte Carlo simulation by sampling $C_{n}^2(0)$ values from the \gls{cdf} shown in Fig.~\ref{fig:cn2_cdf}, generating $1000$ realizations for each acquisition. A threshold of $C_{n, \text{max}}^2(0) = 2 \times 10^{-14},\text{m}^{-2/3}$ is applied to evaluate these cases.

The \gls{eo} satellite captures a given number of \gls{fps} with a resolution of $600\times 600$\,pixels and an average \gls{gsd} of $0.5$\,m, which might be affected by atmospheric turbulence. On average, the size of the semantic data for ship detection (i.e., ships and bounding boxes) is $336$ bits per image. Executing YOLOv8 for this task requires $W_\text{YOLO}=79.1$ G\glspl{flop} per image. 

We consider both \gls{cpu} and \gls{gpu} architectures for the edge satellites. Specifically, three different configurations are considered, where each satellite has either an 8-core CPU running at up to $1.8$\,GHz with a power consumption of $6$\,W, or an NVIDIA Jetson Orin GPU, which can be either a Nano Super or AGX, 
with a maximum power consumption of $15$ and $60$\,W, respectively. The \gls{gs} has a 64-core 
CPU operating at a maximum frequency of $2.6$\,GHz  
with a maximum power consumption of $280$\,W.  

For the considered constellation, average image size $D_\text{img}=788.513$~kbits and compression ratio $\rho=2346$. Under these conditions, the communication latency due to the transmission of both uncompressed $T_\text{comm}^{(u)}(s,p,D_\text{img})$ and compressed data $T_\text{comm}^{(c)}(p,e,D_\text{img})$ can be closely approximated through the propagation latency per intra-plane \gls{isl} for an orbital plane with $N$ satellites in the Edge Layer deployed at an altitude $h_e$:
\begin{equation}
    T_\text{prop}(N, h_e)=\frac{d_\text{ISL}(N,h_e)}{c}=\frac{2(R_E+h_e)\sin\left(\pi/N\right)}{c},
\end{equation} 
where $R_E$ is the radius of the Earth. The later gives $3.1$~ms for the selected $N=23$ and $h_e=617$\,km, which is much larger than the time to transmit one image per \gls{isl} $D_\text{img}/R_\text{ISL}\leq0.1$~ms. Building on this, the upper bound in propagation delay for a path between satellites in the Edge Layer following a shortest-path routing algorithm is $\left\lfloor N/2\right\rfloor T_\text{prop}(N, h_e)$, which gives $34.1$~ms for $N=23$. Therefore, during optimization, we approximate $T_\text{comm}^{(c)}(p,e,D)$, defined in~\eqref{eq:comm_latency}, with the latter upper bound of the propagation latency. Note that the upper bound of propagation latency for any value of $N$ is achieved as $N\to\infty$ and is $2(R_E+h_e)/c$, which gives $36$\,ms for the selected $h_e$.

The summary of main parameters and values can be found in Table~\ref{tab:parameters}. 

\begin{table*}[!t]
\centering
\renewcommand{\arraystretch}{1.22}
\caption{Parameters and settings.}

\begin{tabularx}{\textwidth}{@{}X@{}}
\toprule
{
\begin{minipage}{\linewidth}
\centering
\begin{tabularx}{\linewidth}{@{}Xcccc@{}}
\textbf{Parameter} & \multicolumn{4}{c}{\textbf{Settings}}\\
\cmidrule(lr){2-5}
& \textbf{Cloud CPU} & \textbf{Satellite CPU} & \multicolumn{2}{c}{\textbf{Satellite GPU}}\\
\textbf{Processing architectures} & AMD EPYC 7H12 & Snapdragon 855 & Nano & AGX\\
\midrule
Number of cores $N_\text{cores}$ & 64 & 8 & 1024 & 2048\\
Max.\ power consumption $P_\text{max}$ [W] & 280 & 6 & 25 & 60\\
Max.\ clock frequency $f_\text{max}$ [GHz] & 2.6 & 1.8 & 1.02 & 1.3\\
Number of FLOPs per Hz $N_\text{FLOPs}$ & 32 & 32 & 2 & 2\\
Processing inefficiency $\mu_C$ & 1.079 & 1.079 & 1.071 & 1.122\\
Mean synchronization time $\mu_\text{Tsync}$ [ms] & 0 & 0 & 17.48 & 14.14\\
Max. processing speed $1/\mu_T^{(p)}(f_\text{max}^{(p)})$ [FPS] & $62.377$ & $5.398$&$17.231$&$32.45$\\
\end{tabularx}
\end{minipage}
}
\\
{
\begin{minipage}{\linewidth}
\begin{tabularx}{\linewidth}{@{}l>{\centering\arraybackslash}X@{}}
\midrule
\textbf{Data acquisition} \\\midrule
No. of images in a frame, for training, for evaluation ($N_\text{img}$, $N_\text{img}^\text{train}$, $N_\text{img}^\text{eval}$) & 2601, 1668, 50 images\\
Image resolution $w_\text{img}\times h_\text{img}$ & 600×600 pixels\\
Average frame, image, ship 
{($D_\text{frame}$, $D_\text{img}$, $D_\text{ship}$)}
& 2.05$\cdot10^{6}$, 788.513, 6.913\,
kbits\\
$\text{GSD}$, $\text{GSD}_{\text{nadir}}$ & 0.43, 0.31 m\\
\gls{sth}&1600 s\\
Time slot $T_{\text{slot}}$ & 10 s\\
Power for camera maneuvering $P_{\text{man}}$& 2 W\\
Total energy budget for image acquisition $E_{\text{max}}$& 1000 J\\
\midrule
\textbf{Communication network} \\\midrule
Number of orbital planes $N_p$ & 1\\
Edge satellites per plane $N$ & 23\\
Altitude $h_o, h_e$ & 617, 617 km\\
Inclination $i_o, i_e$ & 98.6, 98.6$^\circ$\\
Edge Layer ISL data rate $R_\text{ISL}$ & 10 Gbps\\
Edge Layer ISL transmit power $P_\text{ISL}$ & 60 W\\
Satellite‑to‑ground transmit power $P_\text{DL}$ & 10 W\\
Downlink bandwidth $B$ & 500 MHz\\
Noise power $\sigma^2_\text{dB}$ & –119.32 dBW\\
Total downlink antenna gain $G_\text{DL}$ & 66.33 dBi\\
\end{tabularx}
\end{minipage}
}
\\
\bottomrule
\end{tabularx}
\label{tab:parameters}
\end{table*}

\subsection{Observation scheduling}
To evaluate observation scheduling performance, we present two sets of results. The first demonstrates how an effective optimization of the \gls{aeossp} leads to high-quality data acquisition. The second assesses the impact of atmospheric turbulence. 

First, we evaluate the scheduling performance in an idealized environment, disregarding atmospheric turbulence effects. The \gls{aeossp} is solved for a single \gls{aeos}, problem instances of $\{80, 100, 120, 140\}$ targets (i.e., areas with vessels), and a $\text{STH} \approx 1600\,\text{s}$. Our proposed solution is compared against schedules generated by two conventional benchmark algorithms: \gls{fifo} and \gls{ga}. For the \gls{ga}, we utilize the implementation provided by the \textit{pymoo} library \cite{pymoo}, extended with custom crossover and mutation operators to ensure schedule feasibility. Specifically, the population size is set to 20, the number of generations to 100, and both crossover and mutation probabilities are fixed at 0.2.

\begin{figure*}[t]
    \centering
    \subfloat[]{
        \includegraphics[width=0.48\textwidth]{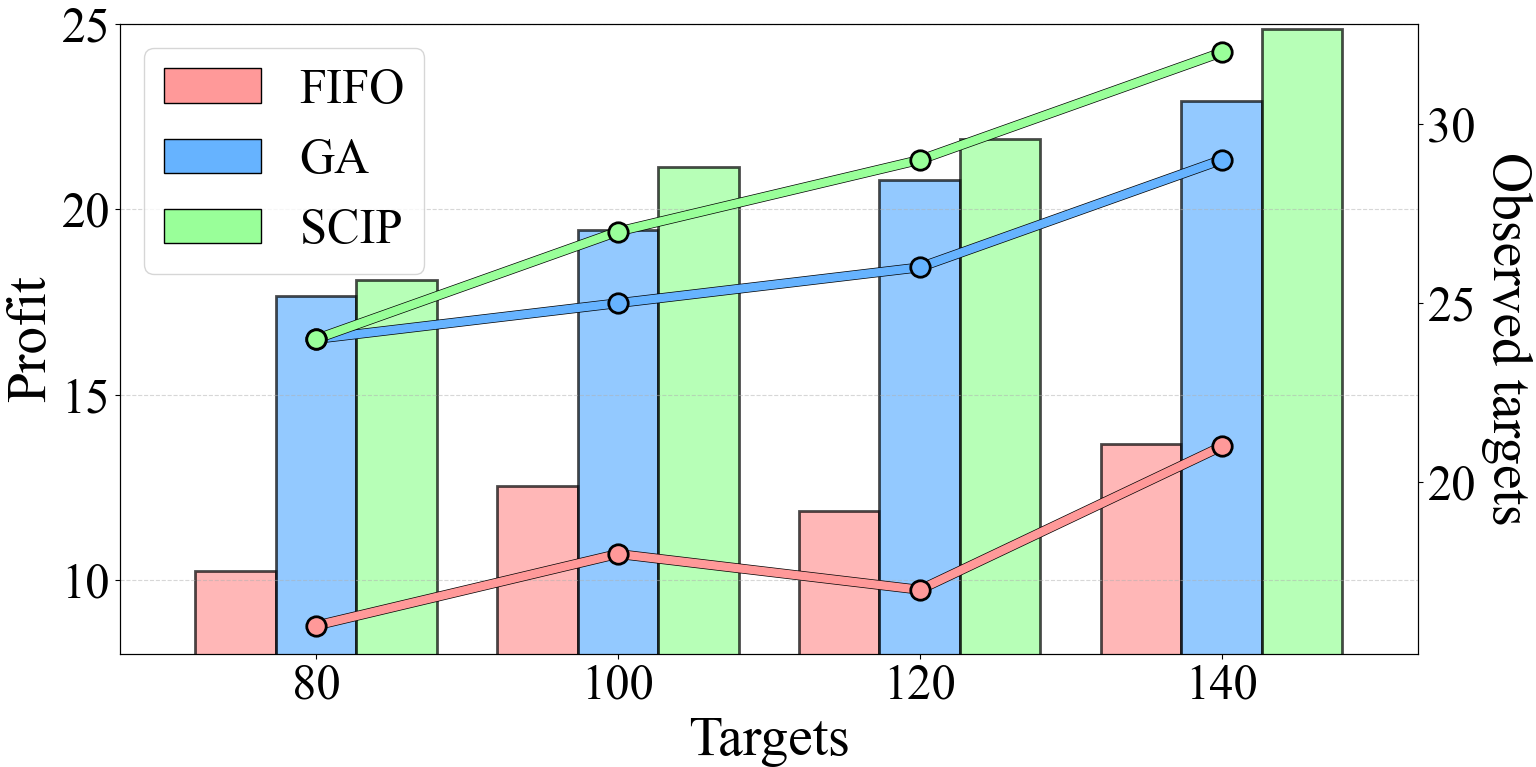}
        \label{fig:observation_profit_draft}}
    \hfill
    \subfloat[]{
        \centering
        \includegraphics[width=0.48\textwidth]{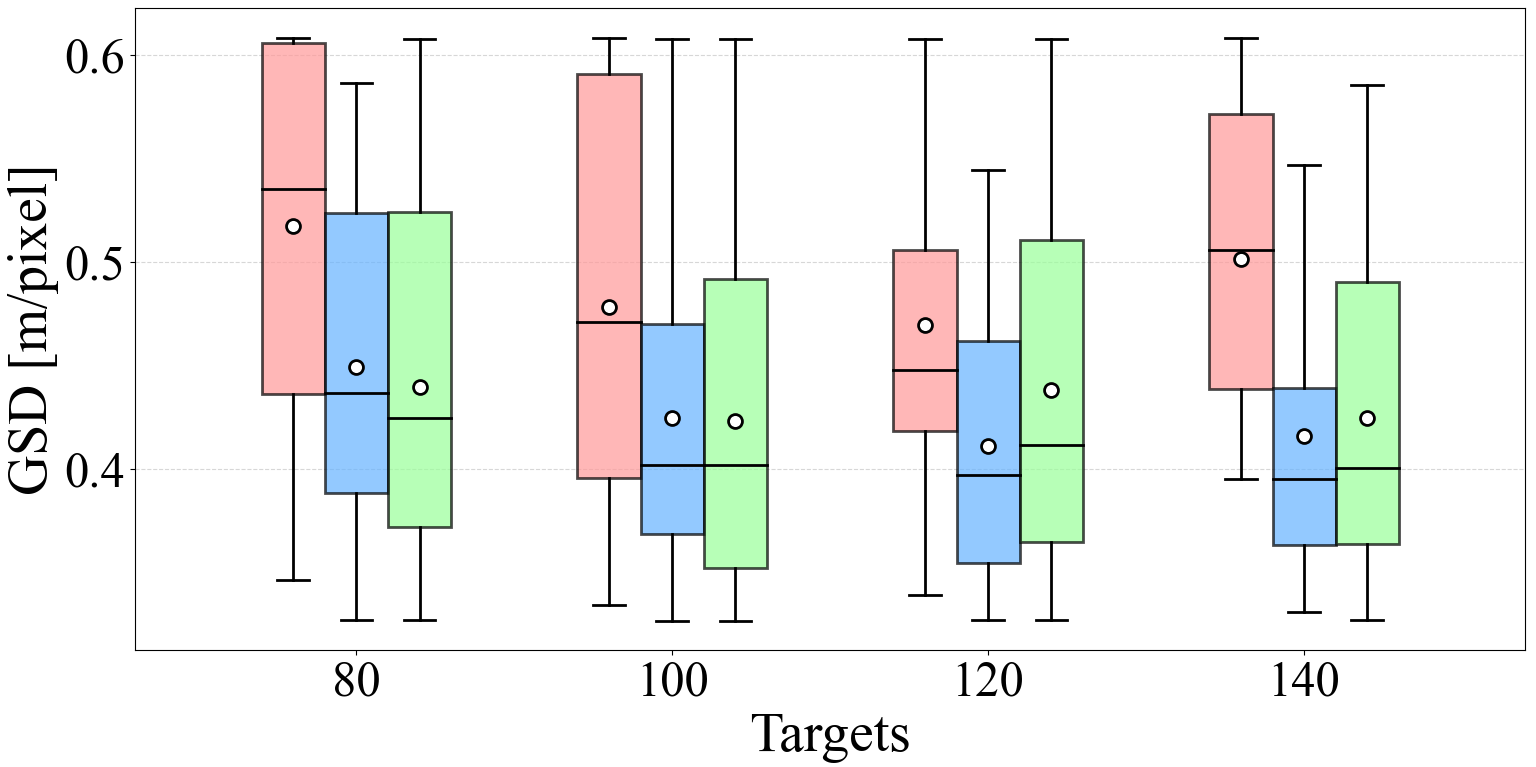}
        \label{fig:boxplots_gsd}}
    \caption{Performance results for the turbulence-free scenario. (a) Observation profit (left) and number of observed targets (right). (b) \gls{gsd} distribution of the acquired data.}
    \label{fig:observation_results_draft}
    \vspace{-0.4cm}
\end{figure*}

As shown in Fig.~\ref{fig:observation_results_draft}, a higher observation profit correlates with a larger number of observed targets, and acquisitions with lower \gls{gsd}, i.e., better image spatial resolution. In turn, improved spatial resolution enhances the accuracy of the semantic extraction algorithm, YOLOv8 in this study. To quantify this impact, evaluating the algorithm's performance in terms of recall (the ratio of correctly detected objects to total actual objects) reveals a drop from 87\% to 73\% when images are downsampled to simulate a \gls{gsd} twice the original. Thus, effective observation scheduling, achieved through a precise solution of the \gls{aeossp}, significantly boosts overall system performance. Specifically, the exact solution given by SCIP yields an average observation profit improvement of 77.86\% and 6.25\% compared to \gls{fifo} and \gls{ga}, respectively. Furthermore, as illustrated in Fig.~\ref{fig:boxplots_gsd}, the proposed schedule attains an average \gls{gsd} of approximately $0.43\,\text{m/pixel}$. This matches the characteristics of one of the reference datasets \cite{ijgi11080445}; consequently, this value will be assumed for the subsequent processing scheduling section. The output of this scheduling is the set of $\text{OTW}_{s,t,o,w}$, $\mathrm{OTW}^\star$, and corresponding times $\tau_{s_k,t_k,o_k,w_k}$ at which the edge nodes will receive new frames to be processed. 

\begin{figure*}[t]
\centering
\includegraphics[scale=0.25]{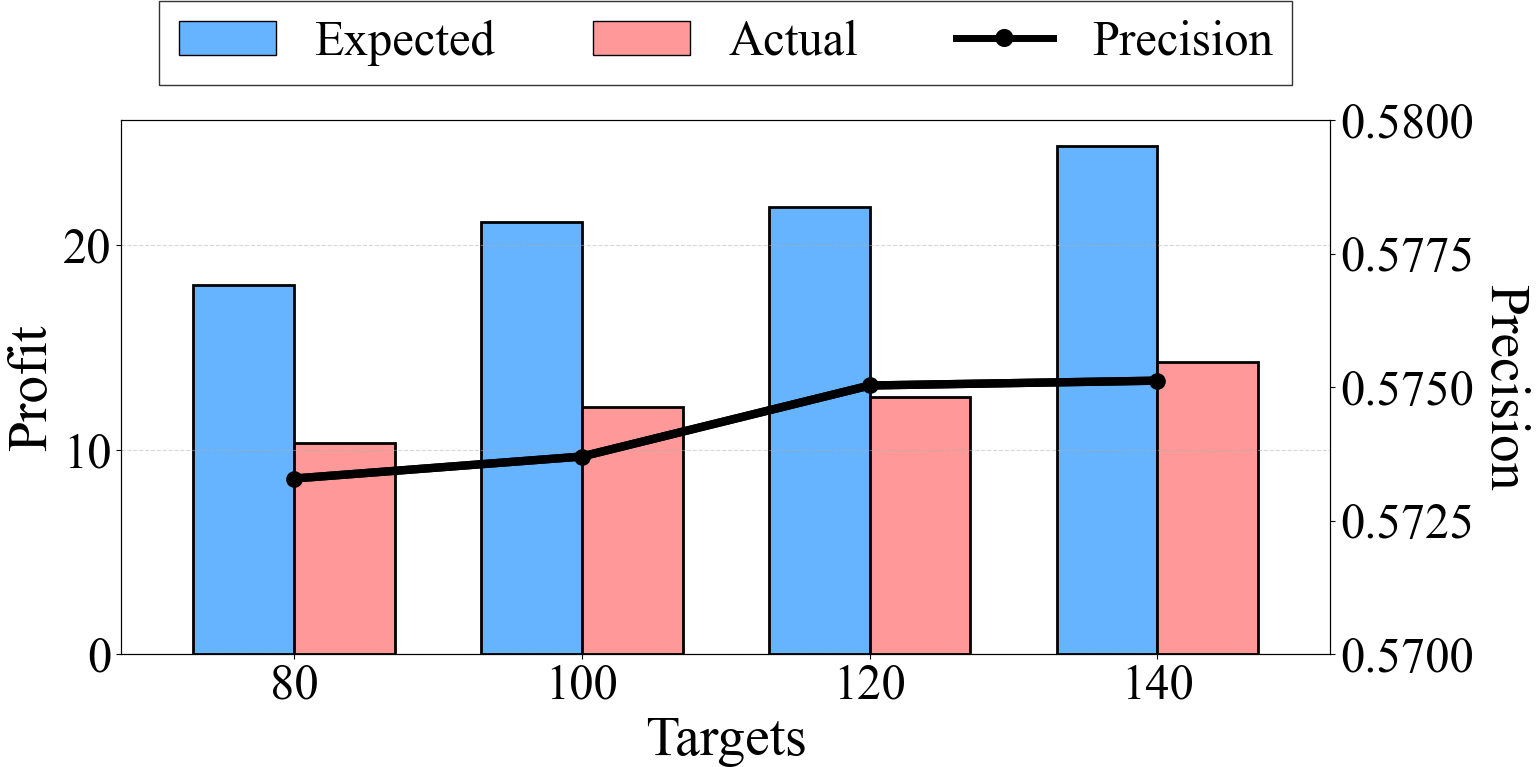}
\caption{Impact of atmospheric turbulence ($C_{n, \text{max}}^2(0) = 2 \times 10^{-14},\text{m}^{-2/3}$) on the actual collected observation profit (left) and average schedule precision defined as the ratio of observations meeting quality requirements (right).}
\vspace{-0.4 cm}\label{fig:observation_profit_turbulence}
\end{figure*}

When the atmospheric turbulence is considered, with a threshold $C_{n, \text{max}}^2(0) = 2 \times 10^{-14},\text{m}^{-2/3}$, Fig.~\ref{fig:observation_profit_turbulence} illustrates the divergence between the expected and actual average collected profit.  .Additionally, the figure displays the average schedule precision, defined as the portion of observations meeting quality requirements. Two key conclusions emerge: (1) the actual profit is significantly lower than expected; and (2) the proportion of acquisitions failing to meet quality standards is substantial. It should be noted that, although the volume of data declines in a similar way to observation profit, it does not do so in exactly the same way, as the value of each individual acquisition is different. This difference would probably be accentuated in conventional scenarios where each observation target is associated with a different priority. Therefore, accounting for atmospheric turbulence is essential for optimal resource allocation. It avoids wasted energy during image acquisition and attitude maneuvers, while also reducing the load on data processing and transmission, ultimately leading to more efficient satellite resource management.

\begin{table}[t]
    \centering
    \caption{Observation performance.}
    \label{tab:observation_performance_turbulence}
    \renewcommand\theadfont{\bfseries} 
    
    \begin{tabular}{ccccc}
        \toprule 
        \thead{Number of \\ targets} & 
        \thead{Total \\ profit} & 
        \thead{Successfully observed \\ targets [\%]} & 
        \thead{Average observations \\ per target} & 
        \thead{Rescheduled \\ targets [\%]} \\
        
        \midrule
            \textbf{150} & 109.16 & 90 & 1.49 &  36.5 \\
        \midrule
            \textbf{200} & 148.06 & 84.5 & 1.46 & 38.4 \\
        \midrule
            \textbf{250} & 177.18 & 86.8 & 1.5 & 36.3 \\
        \midrule
            \textbf{300} & 205.15 & 86 & 1.5 & 33.6 \\
        \bottomrule
    \end{tabular}
\end{table}

Without prior knowledge of the $C_n^2$ value during the schedule generation, we next evaluate the impact of targets observed under unfavorable conditions that are deemed invalid and re-queued for subsequent scheduling processes. For this simulation, we consider a constellation of four \glspl{aeos} arranged in a Walker Star topology, with each satellite in its own orbital plane. The \glspl{aeos} are tasked with observing target sets of sizes $\{150, 200, 250, 300\}$ over 10 consecutive \glspl{sth}, where each \gls{sth} corresponds to one orbital period ($\approx 97$ minutes). If a target is observed satisfying $C_n^2(0) \leq C_{n, \text{max}}^2(0)$, it is considered successfully acquired, and its corresponding candidate \glspl{otw} are removed from subsequent \glspl{sth}. Conversely, if such condition is not met, the target is retained in the problem instance for future scheduling. Consequently, certain targets may undergo multiple observation attempts before a high-quality image is obtained. Table~\ref{tab:observation_performance_turbulence} summarizes the results of the simulation, specifically: the actual collected profit (accounting for turbulence), the percentage of successfully observed targets, the average number of observations per target, and the percentage of rescheduled targets. It is observed that each target undergoes an average of $1.5$ observation attempts, with $33.6\%$ to $38.6\%$ of targets requiring at least two attempts due to turbulence effects. Despite these rescheduling efforts, full coverage is not attained in every instance, with success rates ranging from $84.5\%$ to $90\%$. Finally, as with the single-satellite experiment, instances with a larger number of targets yield a higher total profit.

\subsection{Processing scheduling}
To optimize the processing of the collected data at the Edge Layer, we first characterize the execution time of the task of interest, i.e. semantic extraction using YOLOv8, on the considered computing architectures. Then, we illustrate the performance of distributed edge processing for the different computing architectures in an isolated and controlled snapshot, where the task parameters and network topology have been fixed. Finally, we showcase the performance of the optimization in a realistic scenario with scheduled observations throughout the satellite orbit.

As a starting point, we perform a statistical analysis of the execution time for single images and for the available clock frequencies. We then employ a \gls{bsp} model to extend the mean execution time to other frequencies and the \gls{clt} to derive a statistical model for the execution time of an arbitrary number of images.

\begin{figure}[t]
    \centering
    \includegraphics{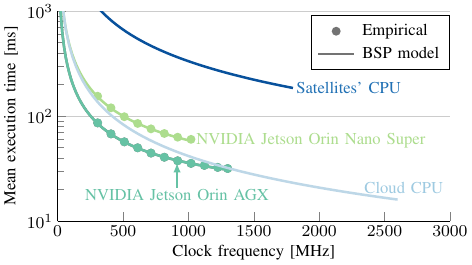}
    \caption{Estimated mean execution time of YOLOv8 on individual images with the considered computing architectures obtained empirically for all the operating clock frequencies $f_p$ (markers) and using the least-squares regression for the \gls{bsp} model parameters (solid lines).}
    \label{fig:exec_time_bsp_model}
\end{figure}

Specifically, we first obtained the mean execution time $\mu_T^{(p)}(f_p)$ of YOLOv8 on individual images executed on the GPU NVIDIA Jetson Orin platforms. For this, the mean execution times were estimated for each available operating frequency $f_p$ using all the images in the dataset. Then, to estimate the mean execution time for all $f^{(p)}\in (0,f_\text{max}^{(p)})$, we performed a least-squares regression for the parameters of the \gls{bsp} model $\mu_C^{(p)}$ and $\mu_\text{Tsync}^{(p)}$. Table~\ref{tab:parameters} lists the operational parameters for the GPUs and the estimated parameters for the \gls{bsp} model. Recall that the complexity of YOLOv8 is $W_\text{YOLO}=79.1$\,G\glspl{flop}.  
The results of the estimated mean execution times are shown in Fig.~\ref{fig:exec_time_bsp_model}, where the mean execution times obtained empirically are shown with markers and the solid lines show the estimation using the \gls{bsp} model with least-squares regression. Fig.~\ref{fig:exec_time_bsp_model} also shows the estimated mean execution time using the \gls{bsp} models for the CPUs at the satellite and \gls{gs}, which were adapted following same procedure on similar CPU architectures and scaled to the number of cores $N_\text{cores}^{(p)}$  and CPU clock frequencies of the models considered in Table~\ref{tab:parameters}. Clearly, the \gls{bsp} model provides a nearly perfect fit to the mean execution times obtained in our experiments, which validates its suitability for the numerical optimization of the processing at the Edge Layer.

\begin{figure}[t]
    \includegraphics[width=\textwidth]{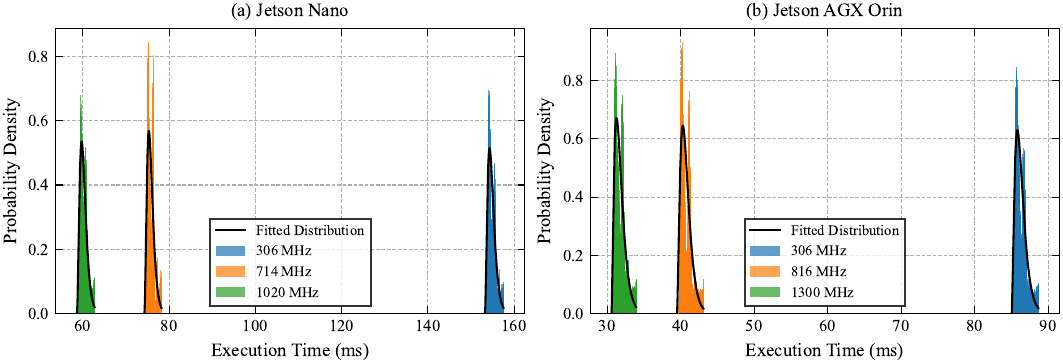}
\caption{Probability density of YOLOv8 execution time on the NVIDIA Jetson Orin (a) Nano Super and (b) AGX  platforms at selected GPU clock frequencies. Solid lines represent fitted Gamma distributions, indicating the execution time distribution follows $T_\text{proc}^{(p)}(f_p)\sim \text{Gamma}(\alpha(f_p),\theta(f_p))$.}
\label{fig:exec_time_distribution}
\end{figure}

Next, we aim to characterize the randomness of the execution time on the GPU architectures. Fig.~\ref{fig:exec_time_distribution} shows the results obtained for three clock frequencies $f_p^{(p)}$ available at both Nano Super and AGX models, where we observe that the execution time of YOLOv8 individual images executed on the GPU NVIDIA Jetson Orin platforms $T_\text{proc}^{(p)}\left(f_p\right)$ at the selected clock frequencies follows a $\text{Gamma}\left(\alpha(f_p),\theta(f_p)\right)$ distribution with mean $\mu_T^{(p)}(f_p)=\alpha(f_p)\theta(f_p)$ and variance $\text{var}_T^{(p)}=\alpha(f_p)\theta(f_p)^2$. Interestingly, we observed that the execution time of YOLOv8 on images with a high object density can exhibit a bimodal distribution, primarily due to \gls{nms} post-processing. For simplicity, our fitted Gamma distribution captures the dominant mode, which corresponds to the most typical scenario in maritime observation conditions.


\begin{figure}[!t]
    \centering

\subfloat[Empirical and fitted Gamma parameters]{
    \includegraphics[width=0.42\linewidth]{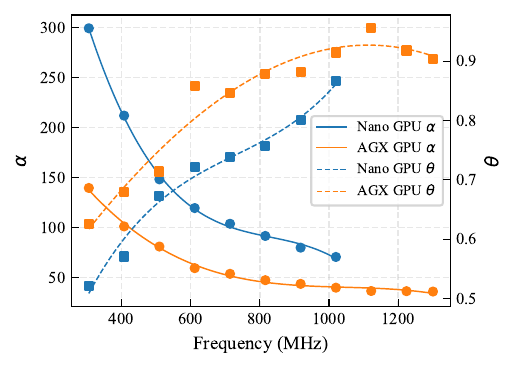}\label{fig:gamma_regression}}\hfil
\subfloat[Quantiles for \gls{clt} approximation]{
    \includegraphics[width=0.48\linewidth]{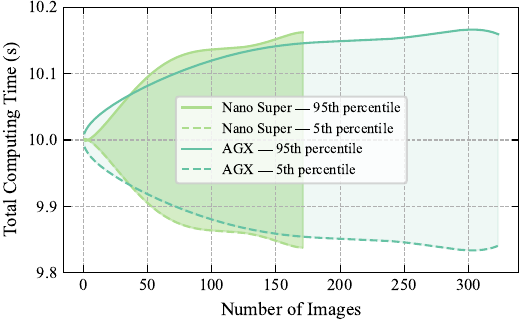}\label{fig:clt_approx}}
    \caption{(a) Empirical and fitted Gamma distribution parameters $\alpha(f_p)$ and $\theta(f_p)$ for YOLOv8 execution time on the NVIDIA Jetson Orin GPU architectures and (b) 5-th and 95-th quantiles for the GPU computing time for a time slot duration of $T_\text{slot}=N_\text{img}\mu_T(f_p^*)=10$ seconds for the NVIDIA Nano Super and AGX models with \gls{clt} approximation $\set{N}\left(T_\text{slot},N_\text{img}\alpha(f_p)\theta(f_p)^2\right)$. }\vspace{-0.4cm}
\end{figure}
As with the mean execution time, the parameters $\alpha(f_p)$ and $\theta(f_p)$ for the  execution time of YOLOv8 were estimated for each available operating frequency $f_p$ using all the images in the dataset. The resulting distribution parameters are shown with markers in Fig.~\ref{fig:gamma_regression}, where both platforms exhibit similar qualitative behavior: the shape parameter $\alpha(f_p)$ decreases with frequency, while the scale parameter $\theta(f_p)$ increases. To enable analytical treatment in the optimization framework, we approximate the frequency-dependent distribution parameters using third-order polynomial regression $\sum_{i=0}^3c_i f_p^i$
and obtaining the approximated parameters $\hat{\alpha}(f_p)$ and $\hat{\theta}(f_p)$ 
shown with line plots 
in Fig.~\ref{fig:gamma_regression}.

Building on these models, we approximate the \gls{rv} for the execution time $T^{(p)}$ for all $f^{(p)}\in (0,f_\text{max}^{(p)})$ and $N_\text{img}\in\mathbb{R}^+$  by applying the \gls{clt} with the mean execution time per image obtained from the \gls{bsp} model $\mu_T^{(p)}$ and the variance of the Gamma distribution $\text{var}(f_p)=\hat{\alpha}(f_p)\left(\hat{\theta}(f_p)\right)^2$. Therefore, we approximate the execution time of YOLOv8 on $N_\text{img}$ images using GPU $p$ operating at frequency $f_p$ as
\begin{IEEEeqnarray}{rCl}
    T_\text{proc}^{(p)}(f_p, N_\text{img})=\sum_{N_\text{img}} T_\text{proc}^{(p)}\left(f_p\right) &\sim& \text{Gamma}\big(N_\text{img}\alpha(f_p),\theta(f_p)\big)\\
    &\approx& \mathcal{N}\left(N_\text{img}\mu_T^{(p)}(f_p), N_\text{img}\text{var}(f_p)\right) \text{ as } N_\text{img}\to\infty
\end{IEEEeqnarray}
Fig.~\ref{fig:clt_approx} shows the approximated 5th and 95th percentiles of the execution time for $N_\text{img}$ in the GPU platforms for a target $N_\text{img}\mu_T^{(p)}(f_p)=10$\,seconds. As it can be seen, the 5th and 95th quantiles deviate from the mean by at most $1.8$\% for all the considered values of $N_\text{img}$. This 
small deviation from the mean, in combination with a negligible delay due to gather extremely compressed data towards the downlink satellite $T_\text{comm}^{(c)}(p,e,D_\text{img})\leq 36$~ms, simplifies the scheduling of the downlink, as it implies that, even when $N_\text{img}\mu_T^{(p)}=T_\text{slot}$ and the downlink is scheduled $k+2$ time slots after a set of images is captured, a large fraction of the slot (i.e., above $98$\%) will be available to transmit the processed images to the \gls{gs}. 

\begin{figure}[t]
\centering
\subfloat[Satellite's CPU]{\includegraphics{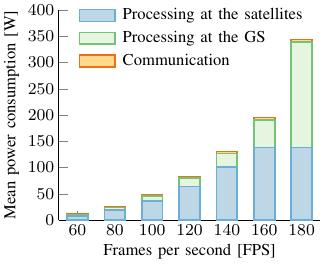}\label{fig:power_cons_cpu}}\hspace{-0.5em}
\subfloat[NVIDIA Jetson Orin Nano Super]{\includegraphics{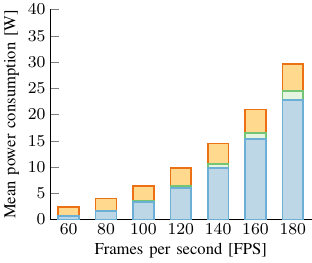}\label{fig:power_cons_nano}}\hspace{-0.5em}
\subfloat[NVIDIA Jetson Orin AGX]{\includegraphics{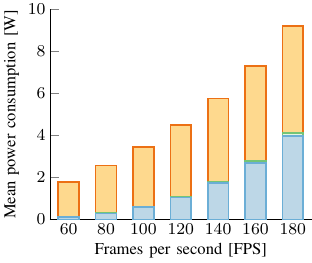}\label{fig:power_cons_agx}}
\caption{Power consumption for the three considered computing architectures at the satellites for 60 to 180 \gls{fps}. Communication happens at the edge of coverage and the source satellite is 5 hops away from the GS.}
\label{fig:power_cons_worst_case_5hops}
\end{figure}

With the experimental characterization of the execution time, we can now address the optimization problem $\mathrm{P}_{\text{sch}}$. In Fig.~\ref{fig:power_cons_worst_case_5hops}, we plot the mean power consumption versus the computing load for a single time slot for the three considered computing architectures at the satellites. The load is given in \gls{fps}, ranging between $60$ to $180$, and the power consumption is divided between the processing at the satellites, the ground station, and communication. The network topology is fixed, considering an \gls{eo} satellite five hops away from the \gls{gs}, located at the edge of coverage of the serving satellite, leading to a data rate of $R_k=2.3$~Gbps for the feeder link.

The power consumption for directly downloading the images and processing at the \gls{gs} is not included in the plots, as the maximum supported load is only $63$ \gls{fps} and leads to a mean power consumption over the time slot of $250$ W.
Furthermore, note that the mean power consumption over the time slot is nearly $10\times$ greater with the satellite's CPU, shown in Fig.~\ref{fig:power_cons_cpu}, than with the Nano Super GPU, shown in Fig.~\ref{fig:power_cons_nano}. In turn, the mean power consumption of the latter is between $2\times$ and $3\times$ greater than with the AGX GPU, shown in Fig.~\ref{fig:power_cons_agx}. Thus, processing the images with the AGX GPU is almost $30\times$ more power efficient than with the satellite's CPU. Furthermore, we observed that using the on-board computing architectures at the Edge Layer is more power efficient than offloading the computations to the \gls{gs} when the computational load is sufficiently light. Then, as Fig.~\ref{fig:power_cons_cpu} and Fig.~\ref{fig:power_cons_nano} show, the use of the CPU at the \gls{gs} increases with the computing load. In particular, the CPU of all the satellites in the Edge Layer must operate at maximum $f_p$ to process $160$~\gls{fps} and higher loads. Consequently, the power consumption due to processing at the satellites is constant for both $160$ and $180$ \gls{fps} in Fig.~\ref{fig:power_cons_cpu}, and this computing architecture cannot support $200$~\gls{fps}. On the other hand, the power consumption at the \gls{gs}'s CPU with the GPU architectures at the Edge Layer is small compared to the total, as the configuration with the Nano Super GPU can support up to $460$~\gls{fps} and the configuration with the AGX GPU can support up to $820$ \gls{fps}.

Next, we evaluate the real-time energy consumption of the optimal processing scheduling in a realistic scenario where the \gls{aeossp} is solved for observing $140$ target areas over an episode of $2000$~seconds and each observation results in $2601$ images captured at the \gls{aeos}. The total number of observations in the period is $32$, which leads to an average of one observation every $62.5$\,seconds. For the ground segment, we consider the $26$ KSAT \glspl{gs}~\cite{KSATGroundNetwork} and, for each time slot $k$ of duration $T_\text{slot}=10$\,seconds, the \gls{gs} that offers the highest downlink rate at the scheduled time slot $k+2$ is selected as destination \gls{gs}.

\begin{figure}[t]
\centering
\subfloat[NVIDIA Jetson Orin Nano Super]{\includegraphics{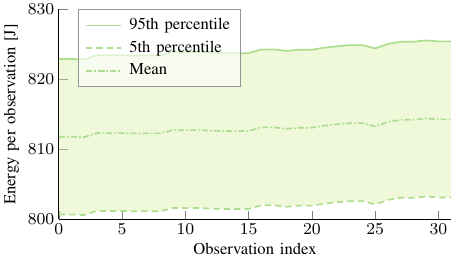}}\hspace{-0em}
\subfloat[NVIDIA Jetson Orin AGX]{\includegraphics{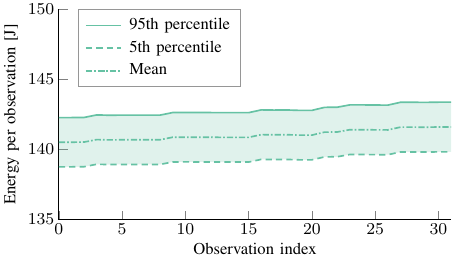}}
\caption{Mean, 5th, and 95th quantiles for the optimal energy consumption of semantic extraction at the satellites with NVIDIA Jetson (a) Nano Super and (b) AGX. A total of 32 observations were performed with the moving satellites.
}
\label{fig:moving_e_cons}
\end{figure}

Fig.~\ref{fig:moving_e_cons} shows the mean, 5th, and 95th, quantiles of the energy consumption per observation with the optimal processing scheduling for the two GPU architectures: NVIDIA Jetson Oring Nano Super and AGX. Clearly, the energy per observation with the Nano Super GPU is more than $5\times$ greater than with the AGX GPU, which is due to the high computing load of $260$ \gls{fps} at each observation. Another clear trend is the increase in energy consumption with the number of observations, which is a consequence of the \gls{aeos} moving away from a group of \glspl{gs} in Svalvard, Troms\o, and Vard\o, Norway, which offer the highest downlink rates throughout the considered period. 
Moreover, the 5th and 95th quantiles of the energy consumption per observation deviate between $1.2$ and $1.4$\% from the mean, which is due to the relatively small deviation of these quantiles for the execution time of the YOLOv8 algorithm.

        


\begin{figure}[!h]
\centering
\includegraphics{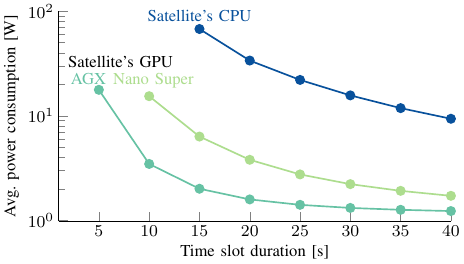}
\caption{Optimal average power consumption for EO with the three considered satellite edge computing architectures and for time slot durations between $5$ and $40$ seconds.}
\label{fig:avg_power_moving_sat}
\end{figure}

Finally, we plot the average power consumption (in logarithmic scale) throughout the observation period of $2000$ seconds for different time slots durations $T_\text{slot}=\{5, 10, \dotsc, 40\}$\,seconds and for the three considered computing architectures in Fig.~\ref{fig:avg_power_moving_sat}. Since the average period between observations is $62.5$ seconds, all the considered values of $T_\text{slot}$ lead to a stable system. However, the minimum period between observations in the optimal schedule is $18$~seconds, which means that all the cases with $T_\text{slot}$ above this value would lead to new observations been taken while computing tasks from previous observations are still being processed.
Among the three computing architectures, only the AGX GPU is able to support time slots of $5$\,seconds, which result in a computing load of $520.2$ \gls{fps} and the CPU architecture can only support time slot durations of $15$\,seconds or higher, leading to computing loads below $174$ \gls{fps}. As in previous results, the AGX GPU is the computing architecture that leads to the lowest average power consumption, but increasing the time slot duration from $20$ to $40$ seconds leads to a minimal decrease in average power consumption from $1.59$ to $1.23$ W, which is mostly required to scatter the data to nearby satellites.

\section{Conclusions} \label{sec:conclusions}
Considering the huge volume of data generated by \gls{eo} missions, we have investigated the potential of leveraging the edge computing power provided by a \gls{leo} satellite constellation to support real-time object localization and detection of vessels, specifically using YOLOv8. The framework is organized into two layers: the Observation Layer, which determines what and how to observe targets; and the Edge Layer, which decides where processing occurs and how results are transmitted to the ground. For the former, atmospheric turbulence impacts both quality and quantity of the images with the observed targets, while the degrees of freedom provided by \glspl{aeos} increase the complexity of the observation scheduling. For the latter, processing is constrained by the pipelined workflow of scattering the captured images across satellites, performing the computations, and transmitting the processed information to ground via the satellite constellation. Overall, this work highlights the strong potential of distributed edge intelligence to improve the efficiency, autonomy, and responsiveness of future satellite-based \gls{eo} systems.

\bibliography{resf}
\bibliographystyle{IEEEtran}




\end{document}